\documentclass[12pt]{article}
\usepackage{amssymb}

\usepackage{graphicx}
\usepackage{amsmath}

\newtheorem{theorem}{Theorem}

\newtheorem{definition}[theorem]{Definition}

\newtheorem{lemma}[theorem]{Lemma}

\newtheorem{remark}[theorem]{Remark}

\input{tcilatex}

\begin{document}

\baselineskip16pt

\begin{titlepage}

\  \\

\vspace{16 mm}

\begin{center}
{\Large \bf On the General Structure of the }\\
\vspace{2mm}
{\Large \bf Non--Abelian Born--Infeld Action}\\

\vspace{3mm}

\end{center}

\vspace{8 mm}

\begin{center}

Lorenzo Cornalba

\vspace{3mm}

{\small  \' Ecole Normale Sup\' erieure} \\
{\small  Paris, France} \\
{\small  cornalba@lpt.ens.fr} \\

\vspace{5mm}

June 2000

\vspace{5mm}

\end{center}

\begin{abstract}

{ \it

We discuss the general structure of the non--abelian Born--Infeld action, together
with all of the $\alpha ^{\prime }$ derivative corrections, in flat $D$--dimensional
space--time. More specifically, we show how the connection between open strings
propagating in background magnetic fields and gauge theories on non--com\-mutative
spaces can be used to constrain the form of the effective action
for the massless modes of open strings at weak coupling. In particular,
we exploit the invariance in form of the effective action under a change of
non--commutativity scale of space--time to derive algebraic equations relating
the various terms in the $\alpha ^{\prime }$ expansion. Moreover, we explicitly
solve these equations in the simple case $D=2$, and we show, in particular,
how to construct 
the minimal invariant derivative extension of the NBI action.

}
\end{abstract}

\end{titlepage}
\newpage \begin{tableofcontents}
\end{tableofcontents}
\pagebreak

\section{Introduction}

The dyna\-mics of massless open string states propagating in flat
space--time can be described, at weak string coupling, by an effective
action 
\begin{equation*}
S(A)\sim \frac{1}{g_{s}}\int \text{\textrm{Tr} }\mathcal{L}(A)
\end{equation*}
which is function of a $U(N)$ gauge potential $A$ and where the lagrangian
density $\mathcal{L}$ is, in general, an expansion in $\alpha ^{\prime }$
written in terms of arbitrary powers of the curvature $F$ and of its
covariant derivatives. The action $S$ has various equivalent
interpretations. First of all, it reproduces, at tree level, the disk
amplitudes computed directly in string theory. Secondly, the equations of
motion derived from $S$ correspond to the condition of conformal invariance
of the open string sigma model with Wilson line interactions \textrm{Tr}$%
P\exp \left( i\int_{\partial \Sigma }A\right) $ on the boundary $\partial
\Sigma $ of the string world-sheet. In fact, it was shown by Tseytlin and
Andreev \cite{TA} that the effective action itself can be identified with
the partition function of the $\sigma $--model. Finally we recall that,
using $T$--duality, the action $S(A)$ also describes the weak coupling
dynamics of $D$-branes of all dimensions.

New progress in the understanding of the general form of the effective
action $S$ has been made in a recent work by Seiberg and Witten. In \cite{SW}
the authors show that, if the open string propagates in a space--time with a
constant background NSNS two--form field $B$, then the dynamics of the
open--string modes can be described in two equivalent ways. Firstly, one can
consider the original action $S$, and add to the curvature $F$ a central
term $B\cdot \mathbf{1}$. Alternatively, one can replace the original $U(N)$
gauge theory by a gauge theory on a non--commutative space, with
non--commutativity parameter $\theta $ related to the background metric $g$
and two--form $B$. The action then takes the same form as the original
action $S$, with the gauge potential $A$ replaced with the gauge potential $%
\widehat{A}$ of a non--commutative gauge theory, and with products of fields
replaced with Moyal products $\star $ with parameter $\theta $. More
conjecturally, it is shown in \cite{SW} that one can in fact choose the
parameter $\theta $ freely by properly adjusting the central term $\Phi
\cdot \mathbf{1}$ added to the non--commutative curvature $\widehat{F}$. For
each value of $\theta $, the action describing the open--string dynamics has
then the same exact form as the original action $S$. The parameter $\theta $
is then a redundant parameter, since different values of $\theta $
correspond to the same underlying physics. On the other hand, the simple
fact that the form of the action at various values of $\theta $ is invariant
imposes severe restrictions on the possible structure of the original action 
$S$. This paper is devoted to the understanding of those restrictions.
Previous work on this subject is contained in \cite{Okawa,C}.

Before describing the results let us remark a basic fact. In order to attack
the problem of invariance, one need not consider the full $U(N)$ theory. In
fact, if one restricts the attention to the $U(1)$ case, but considers
non--zero values of the non--commutativity parameter $\theta $, one is then
considering a theory which is effectively non--abelian. From an algebraic
point of view, the requirements imposed by form--invariance of the action
are identical in the $U(1)$ and $U(N)$ case as soon as $\theta \neq 0$. For
this reason we work throughout in the $U(1)$ case, but the results will, at
the end, be valid in the general $U(N)$ setting. It is really quite
remarkable that the simpler abelian theory contains, in an subtle way
indeed, the complete information about the general non--abelian theory.

Let us now describe the general results of this paper. As just remarked,
they are valid in the general non--abelian $U(N)$ setting. An invariant
action $S$ is given as a linear combination $S=\sum_{i}c_{i}I_{i}$ of basic
actions $I_{i}$, which we call invariant blocks. A single block $I$ has
itself the following structure. First write $I$ as an expansion in $\alpha
^{\prime }$, as 
\begin{equation*}
I\propto \sum_{L\geq P}\left( \alpha ^{\prime }\right) ^{L}I_{L}
\end{equation*}
where we call a term proportional to $\left( \alpha ^{\prime }\right) ^{L}$
a term of level $L$. The lowest level term $I_{P}$ is a pure derivative
term. The precise meaning of this notion will be given later in the paper,
but informally we can say that pure derivative terms are those which are
invariant under addition of a central term to the curvature $F$ (the basic
example is the $F^{2}$ term at level $P=2$). The higher level terms $I_{L}$
are then needed in order to achieve invariance of the full action $I$ under
a change of the parameter $\theta $. The basic result of this paper is to
reduce the question of invariance to a set of algebraic equations relating
the various terms $I_{L}$. In particular, we will show that the requirement
of invariance can be rephrased in terms of four basic algebraic operators 
\begin{equation*}
\Delta \,\ \ \ \ \ \ \ \overline{\Delta }\,\ \ \ \ \ \ \ \ \ \ \ \ \ \delta
\,\ \ \ \ \ \ \ \overline{\delta }
\end{equation*}
which depend on an arbitrary antisymmetric matrix $\Delta _{ab}$ and which
satisfy the basic commutation relations 
\begin{equation}
\lbrack \overline{\Delta },\Delta ]=[\delta ,\overline{\delta }]=2L-\frac{D}{%
2},  \label{I2}
\end{equation}
where $D$ is the dimension of space--time. The various terms $I_{L}$ then
must satisfy the equations 
\begin{eqnarray}
\Delta I_{L} &=&\delta I_{L+1}  \label{I1} \\
\overline{\Delta }I_{L+1} &=&\overline{\delta }I_{L}.  \notag
\end{eqnarray}
A term of lowest level satisfies then $\Delta I_{P}=\overline{\delta }%
I_{P}=0 $.

We also show in this paper how the above equations can be explicitly solved
in $D=2$. This is clearly a toy model, since gauge--bosons in two dimensions
do not have propagating degrees of freedom. On the other hand, the algebraic
equations are perfectly well defined in dimension $2$, and are highly
non--trivial. This model is useful for a variety of reasons. First of all,
one can show that, starting from the lowest level $F^{2}$ term, one can
reconstruct the full BI action, plus a minimal set of derivative corrections
which are required for invariance of the action (we show that derivative
corrections already enter at level $4$). Similarly, one expects that the
invariant block built from $F^{2}$ in general dimension $D$ will be the
minimal derivative extension of the NBI action. Finally, the construction in 
$D=2$ is a first indication of how to solve the equations (\ref{I1}) in the
general case.

We have said that the action $S$ is a linear combination of invariant
blocks. The specific coefficients are not constrained by the methods of this
paper, and should be determined by other means. However let us note that
none of the arguments which follow rely on supersymmetry, and therefore we
expect supersymmetry to impose constraints on the coefficients themselves,
restricting even more the set of allowed forms of the action.

We shall now describe the contents of this paper. First let me note that
section \ref{summ} contains a concise summary of the results, which includes
the main equations in the text. The structure of the paper is as follows. In
section \ref{S2}\ we review the general structure of the action $S$, as can
be inferred from the analysis of scattering amplitudes of gauge bosons. We
then use this knowledge in section \ref{S3} to rewrite the effective action
as a matrix action. In doing so, we review the general form of the
Seiberg--Witten map which relates commutative and non--commutative gauge
potentials, and we introduce the formal algebraic machinery which is
required in the sequel of the paper. We also derive the second equation in (%
\ref{I1}). Section \ref{S4} is then devoted to rewriting the results of
section \ref{S3} in an invariant operator language. The purpose is
two--fold. On one side, the structure of the action becomes more
transparent. Moreover, in this setting, the question of invariance from $%
\theta $ is more easily understood and solved. Section \ref{S5} then applies
the results of section \ref{S4} to the specific problem at hand, and
completes the derivation of the equations (\ref{I1}), together with the
basic commutator (\ref{I2}). Section \ref{S6} is then devoted to the
explicit solution of equations (\ref{I1}) in the case of $D=2$. We conclude
in section \ref{S7} with discussion and comments on open problems for future
research.

The results of this paper require, together with the general discussion, a
considerable number of technical lemmata. We have tried to limit the
technical discussion to a minimum in the main body of the paper, leaving the
precise proofs of many statements to a rather large appendix.

\section{\label{S2}The General Form of the Effective Action}

We consider an open string propagating in flat $D$--dimensional space--time $%
M$ with coordinates $x^{a}$ and with constant metric $g_{ab}$. Throughout
the paper we use units such that 
\begin{equation*}
2\pi \alpha ^{\prime }=1.
\end{equation*}

We concentrate on the physics of the massless $U(N)$ gauge bosons, to lowest
order in the string coupling constant $g_{s}$. The amplitude $\mathcal{A}%
(p^{I})$ for the scattering of $n$ gluons with momenta $p^{I}$ -- with $%
g^{ab}p_{a}^{I}p_{b}^{I}=0\,$\ -- is computed starting from the disk $n$%
--point function of the corresponding vertex operators, cyclically ordered
on the boundary of the string world--sheet, and then by summing over the
cyclically inequivalent orderings. More precisely, one has that 
\begin{equation*}
\mathcal{A}(p^{I})\sim g_{s}^{n-2}\sum_{\substack{ \sigma \text{ cyclically} 
\\ \text{inequivalent}}}\mathcal{F}_{g}(p^{\sigma _{1}},\cdots ,p^{\sigma
_{n}})
\end{equation*}
where $\mathcal{F}_{g}$ depends on the metric $g$ and is invariant under
cyclic permutations of the arguments\footnote{%
We omit explicit reference to polarizations.}.

One can equivalently summarize the information about disk amplitudes by
introducing an effective action $S(A)$, function of a $U(N)$ connection $A$
on $M$, such that the tree level amplitudes of $S$ are equal to the disk
amplitudes $\mathcal{A}$. The general form of the action $S$ is well known
and reads 
\begin{equation}
S(A,g,g_{s})=\frac{\mathrm{Tr}}{g_{s}}\int d^{D}x\,\det {}^{%
{\frac12}%
}g_{ab}\left( 1+\frac{1}{4}F_{ab}F_{cd}g^{ac}g^{bd}+\cdots \right) .
\label{eq1-100}
\end{equation}
We have absorbed any numerical prefactor in the definition of $g_{s}$.
Moreover, the terms hidden in $\cdots $ contain both higher powers of the
field strength and derivative terms\footnote{%
Various facts are known about the terms in (\ref{eq1-100}). First of all, in
the $U(1)$ case, the terms without derivatives resum to the Born--Infeld
lagrangian \cite{L,CN} 
\begin{equation*}
S(A)=\frac{1}{g_{s}}\int d^{D}x\,\det {}^{%
{\frac12}%
}\left( g+F\right) .
\end{equation*}
In the non--abelian case even the non--derivative terms are not completely
known. At order $F^{4}$ the computation can be explicitly carried out in
string theory, and the result is proportional to \cite{T} 
\begin{equation*}
\mathrm{Tr}\left( F_{ab}F_{cb}F_{ad}F_{cd}+\frac{1}{2}%
F_{ab}F_{cb}F_{cd}F_{ad}-\frac{1}{4}F_{ab}F_{ab}F_{cd}F_{cd}-\frac{1}{8}%
F_{ab}F_{cd}F_{ab}F_{cd}\right) .
\end{equation*}
At higher orders in $F$, the more reasonable proposal is a natural extension
of the Born--Infeld action proposed by Tseytlin \cite{T} in terms of a
symmetrized trace prescription 
\begin{equation}
S(A)=\frac{\mathrm{STr}}{g_{s}}\int d^{D}x\,\det {}^{%
{\frac12}%
}\left( g+F\right) .  \label{SymTrace}
\end{equation}
The above prescription not only matches (up to order $F^{4}$) with the
scattering computations in superstring theory, but also matches results for $%
D$--brane actions derived within matrix theory. \cite{TV,Myers}. On the
other hand, it is known that the symmetrized trace prescription is
incomplete at order $F^{6}$. In \cite{HT}, the authors study the spectra of
excitations around diagonal and intersecting D--brane configurations on
tori, and find discrepancies with the prescription (\ref{SymTrace}). The
correction terms at order $\alpha ^{\prime 3}$ have been explicitly computed
in \cite{BB}.
\par
Some derivative corrections are known, both in the $U\left( 1\right) $ case,
as well as in the non--abelian setting. For the $U(1)$ theory, some
derivative terms have been computed \cite{TA}. In particular, for bosonic
open string theory, the authors find terms at order $F^{2}\partial F\partial
F$. Still in \cite{TA}, derivative correction in superstring theory at order 
$F^{2}\partial \partial F\partial \partial F$ are discussed. In \cite{Kit},
the author finds derivative corrections at order $F^{5}$ (and $F^{3}D^{2}F$%
), proportional to $\zeta \left( 3\right) $, by studying $5$--point disk
amplitudes. Finally, in the bosonic theory, there is a known derivative term
at order $F^{3}$ which is proportional to 
\begin{equation*}
\mathrm{Tr}\left( F_{ab}\left[ F_{bc},F_{ca}\right] \right) .
\end{equation*}
}. As a note on conventions, in all that follows actions will always be
written assuming a \textit{Euclidean signature} of the metric.

Let us now introduce a constant background NS--NS two--form field $B_{ab}$.
In the presence of open strings, a constant field $B$ is not pure gauge and
does affect the dynamics of the gauge bosons. In particular, the effects
have been very clearly analyzed in \cite{SW,SSS} and can be summarized as
follows

\begin{itemize}
\item  The momenta of the asymptotic gluon states satisfy a modified
on--shell condition. More precisely, gluons are massless with respect to an
effective open string metric $G=g-B\frac{1}{g}B$ and therefore the
corresponding momenta satisfy $G^{ab}p_{a}p_{b}=0$.

\item  The effective coupling constant is modified to an open string value
of $G_{s}=g_{s}\det {}^{%
{\frac12}%
}G\det {}^{-%
{\frac12}%
}\left( g+B\right) $

\item  Finally, the disk scattering amplitudes are modified by momentum
dependent phase factors. In particular, in terms of the antisymmetric matrix 
$\theta ^{-1}=B-g\frac{1}{B}g$, the amplitudes are given by 
\begin{equation}
\mathcal{A}(p^{I})\sim G_{s}^{n-2}\sum_{\substack{ \sigma \text{ cyclically} 
\\ \text{inequivalent}}}\mathcal{F}_{G}(p^{\sigma _{1}},\cdots ,p^{\sigma
_{n}})e^{-\frac{i}{2}\theta ^{ab}\sum_{I>J}p_{a}^{\sigma _{I}}p_{b}^{\sigma
_{J}}}  \label{eq1-200}
\end{equation}
\end{itemize}

The above dynamics can again be summarized in tree diagrams of a modified
effective action which can be written, starting from (\ref{eq1-100}), in
various different but equivalent ways . Let me briefly review the various
options:

\begin{enumerate}
\item  On one hand, one can follow the usual prescription by starting with
the action (\ref{eq1-100}) and by simply adding to the field strength $F$
the central term $B\cdot \mathbf{1}$.

\item  On the other hand, one can follow the ideas of \cite{SW}. In this
case, one considers the action (\ref{eq1-100}) with the replacements $%
g_{ab}\rightarrow G_{ab}$ and\thinspace $g_{s}\rightarrow G_{s}$. This
correctly reproduces the modified gauge coupling and the modified
mass--shell condition. Phase factors in (\ref{eq1-200}) are reproduced by
reinterpreting the matrix $\theta $ as a non--commutative scale of
space--time, and substituting, in the action (\ref{eq1-100}), ordinary
products of fields with Moyal products in terms of $\theta $.
Correspondingly, the $U(N)$ gauge potential $A$ is now mapped into a gauge
potential 
\begin{equation*}
\widehat{A}=\widehat{A}_{SW}(A,\theta )
\end{equation*}
of a $U(N)$ non--commutative gauge theory, and the gauge group is modified
accordingly. We will call the map $\widehat{A}_{SW}$ the Seiberg--Witten
map.\bigskip

N{\small OTATION}. We have denoted in (\ref{eq1-100}) by $S(A,g,g_{s})$ the
effective action at zero $B$ field, as a function of the gauge potential,
the metric, and the coupling constant. At finite $B$, there are two new
relevant parameters in the description of the action -- a possible central
term added to the curvature, and a possible non--commutativity parameter. In
general, the action will then depend on five parameters 
\begin{equation*}
S(\text{potential, metric, coupling, central term, NC paramter}).
\end{equation*}
Then the equivalence of the descriptions $1$ and\thinspace $\ 2$ above is
just 
\begin{equation*}
S(A,g,g_{s},B,0)=S(\widehat{A},G,G_{s},0,\theta ).
\end{equation*}

\item  Finally, one can follow a naive procedure, which is usually not
considered, but which will be important for our future discussion. In fact,
this procedure is the most natural one if we are given only the information
about the amplitudes (\ref{eq1-200}), without any reference to an underlying
string theory. Specifically, we may wish to reproduce directly the
amplitudes (\ref{eq1-200}) using a standard $U(N)$ theory, without using any
previous information about the theory at zero $B$. Firstly, the kinetic term
of the theory must be 
\begin{equation}
\frac{\mathrm{Tr}}{G_{s}}\int d^{D}x\,\det {}^{%
{\frac12}%
}G_{ab}\left( \frac{1}{4}G^{ac}G^{bd}F_{ab}F_{cd}\right)  \label{bingo}
\end{equation}
in order to reproduce the correct mass--shell condition. We can write the
above equation in a more suggestive form by introducing the matrix $\Gamma
=g+B$ and by first noting that, since $\mathrm{Tr}\left( F\wedge F\right) $
is a total derivative, then 
\begin{equation*}
\int d^{D}x\mathrm{Tr}\left( F_{ab}F_{cd}+F_{ac}F_{db}+F_{ad}F_{bc}\right)
=0.
\end{equation*}
This identity, together with the facts that $2G^{ab}=\Gamma ^{ab}+\Gamma
^{ba}$ and that $G_{s}^{-1}\det {}^{%
{\frac12}%
}G=g_{s}^{-1}\det {}^{%
{\frac12}%
}\Gamma $, can be used to show that the kinetic term (\ref{bingo}) is equal
to 
\begin{equation*}
\frac{\mathrm{Tr}}{g_{s}}\int d^{D}x\,\det {}^{%
{\frac12}%
}\Gamma _{ab}\left( \frac{1}{4}\Gamma ^{ac}\Gamma ^{db}F_{ab}F_{cd}+\frac{1}{%
8}\Gamma ^{ab}F_{ab}\Gamma ^{cd}F_{cd}\right) .
\end{equation*}
The above is nothing but the quadratic term coming from the expansion of the
Born--Infeld action $\sqrt{g+B+F}=\sqrt{\Gamma +F}$. The matrix $\Gamma $,
which is not symmetric, now plays the role of the metric and therefore more
tensor structures are possible (in fact, the expansion of the BI action
includes all powers of $F$, including the odd ones). In some sense, we have
traded the central term $B\cdot \mathbf{1}$ added to the curvature $F$ with
an addition to the metric $g\rightarrow g+B$, by allowing metrics to be
non--symmetric. This invariance is natural from the point of view of the
open string $\sigma $--model 
\begin{equation*}
\int_{\Sigma }g_{ab}\partial X^{a}\overline{\partial }X^{b}+\int_{\Sigma
}B+\int_{\partial \Sigma }A
\end{equation*}
where we see, using $\int_{\partial \Sigma }A=\int_{\Sigma }F$, that only
the combination $g+B+F$ has an invariant meaning. We are then led to
conclude that there is an extension of (\ref{eq1-100}) in the case of a
non--symmetric metric $\Gamma $ of the general form 
\begin{equation}
\frac{\mathrm{Tr}}{g_{s}}\int d^{D}x\,\det {}^{%
{\frac12}%
}\Gamma _{ab}\left[ 1-\frac{1}{2}\Gamma ^{ab}F_{ab}+\cdots \right]
\label{finalform}
\end{equation}
which reproduces the amplitudes (\ref{eq1-200}). Moreover, we claim that all
contractions of indices in $\cdots $ are done with $\Gamma ^{ab}$ (no terms
containing $\Gamma _{ab}$). This fact can be shown starting with (\ref
{eq1-200}). In fact, these amplitudes are just functions of $G^{-1}$ and $%
\theta $, which only depend on $\Gamma ^{ab}$. This shows that the
amplitudes do not depend on $\Gamma _{ab}$. To show the same fact for the
vertices of the action, we must show that the subtractions coming from poles
in the various subchannels also share the same property. The only problems
could come from internal propagators $p^{-2}G_{ab}$. We recall though that 
\cite{SW}, for amplitudes of the form (\ref{eq1-200}), a general tree graph
is computed by first analyzing the graph at $\theta =0$, and then by
multiplying it by a phase factor depending only on external momenta.
Moreover, the graph at $\theta =0$ has all propagators $p^{-2}G_{ab}$
contracted with metrics $G^{ab}$ on the vertices, thus proving that the
subtraction is again just a function of $G^{-1}$ and $\theta $. Using the
general form (\ref{finalform}) of the action we can then give a meaning to
the function $S$ when the metric is not symmetric. We can then summarize the
equality of descriptions $1$ and $3$ by saying that 
\begin{equation*}
S(A,g,g_{s},B,0)=S(A,g+B,g_{s},0,0).
\end{equation*}
\end{enumerate}

We have then seen that we can trade a central term $B\cdot \mathbf{1}$ with
either a non--commutativity parameter $\theta $ or with an addition $%
g\rightarrow g+B$ to the metric. It is natural (following \cite{SW,AAA}) to
conjecture that, in fact, one has a continuous family of possibilities,
parametrized by a central term $\Phi \cdot \mathbf{1}$ and by a \textit{free}
parameter $\theta $. The effective non--symmetric metric $\Gamma $ then
combines with the central term $\Phi $ into the invariant combination $%
\Gamma +\Phi $, which, following again \cite{SW}, is given by 
\begin{equation*}
\frac{1}{\Gamma +\Phi }+\theta =\frac{1}{g+B}.
\end{equation*}
The effective coupling again depends only on the sum $\Gamma +\Phi $, and is
given by 
\begin{equation*}
\frac{1}{G_{s}}\det {}^{%
{\frac12}%
}\left( \Gamma +\Phi \right) =\frac{1}{g_{s}}\det {}^{%
{\frac12}%
}\left( g+B\right) .
\end{equation*}
Finally, the gauge potential is given by the Seiberg--Witten map $\widehat{A}%
=\widehat{A}_{SW}(A,\theta )$. The action $S(\widehat{A},\Gamma ,G_{s},\Phi
,\theta )$ is then independent of $\Phi $ and $\theta $.

\section{\label{S3}The Effective Action as a Matrix Action}

In this section we continue our general analysis of the effective action $S$%
, but we restrict our attention to the $U(1)$ case. As already noted in the
introduction, whenever the non--commutativity scale $\theta $ is non--zero,
the $U(1)$ case contains the physics of the full $U(N)$ theory, and we
therefore lose nothing in concentrating on the effective action for $N=1$.

\subsection{Choosing the central term}

In the previous section we have argued that the effective action describing
the dynamics of gluons in space--time is given by a function $S(\widehat{A}%
,\Gamma ,G_{s},\Phi ,\theta )$ of five arguments -- \textit{i.e.} the gauge
potential $\widehat{A}$, the generalized non--symmetric metric $\Gamma $,
the coupling $G_{s}$, the central term $\Phi $ and the non--commutativity
parameter $\theta $.

The arguments of the action $S$ are not all independent, since physically
different backgrounds are parametrized only by the closed string parameters $%
A,g+B$ and $g_{s}$, which we keep fixed. In fact, the same physical
situation corresponds to a family of different values of the arguments of $S$%
, parameterized by $\Phi $ and $\theta $, which we consider as free
parameters. The remaining variables $\widehat{A},\Gamma $ and $G_{s}$ are
then determined, in terms of the fixed closed string parameters, by the
equations 
\begin{eqnarray}
\widehat{A} &=&\widehat{A}_{SW}(A,\theta )  \notag \\
\frac{1}{\Gamma +\Phi }+\theta &=&\frac{1}{g+B}  \label{eq2-100} \\
\frac{1}{G_{s}}\det {}^{%
{\frac12}%
}\left( \Gamma +\Phi \right) &=&\frac{1}{g_{s}}\det {}^{%
{\frac12}%
}\left( g+B\right) .  \label{eq2-200}
\end{eqnarray}
The action $S$ is then independent of $\Phi $ and $\theta $.

We use this freedom to choose the central term $\Phi $. Throughout the paper
we will denote with $K$ the inverse of $\theta $ 
\begin{equation*}
K=\frac{1}{\theta }.
\end{equation*}
Using the independence of $S$ on $\Phi $, we set 
\begin{equation*}
\Phi =-K.
\end{equation*}
The only free parameter is then the non--commutativity scale $\theta $.

Equation (\ref{eq2-100}) can be easily rewritten in terms of the combination 
\begin{equation*}
\gamma =g+B-K
\end{equation*}
and reads 
\begin{equation}
\Gamma =-K\frac{1}{\gamma }K.  \label{eq2-300}
\end{equation}
Finally equation (\ref{eq2-200}), which determines the coupling, becomes 
\begin{equation}
\frac{1}{G_{s}}\det {}^{%
{\frac12}%
}\Gamma =\frac{1}{g_{s}}\det {}^{%
{\frac12}%
}K.  \label{eq2-400}
\end{equation}

\subsection{\label{ama}A matrix action}

Let us now consider the expression for the field strength $\widehat{F}$ and
its covariant derivatives. We start by introducing the coordinate functions 
\begin{equation*}
x_{a}=K_{ab}x^{b}
\end{equation*}
and the combinations 
\begin{eqnarray*}
\lambda _{a} &=&x_{a}+\widehat{A}_{a} \\
\lambda ^{a} &=&\theta ^{ab}\lambda _{b}=x^{a}+\theta ^{ab}\widehat{A}_{b}.
\end{eqnarray*}
Note that we raise and lower indices with the matrix $\theta ,K$. Using the
simple fact that, for any function $f$, 
\begin{equation*}
\partial _{a}f=-i[x_{a},f],
\end{equation*}
we quickly see that the commutator $-i[\lambda _{a},\lambda _{b}]$ is given
by 
\begin{eqnarray*}
-i[\lambda _{a},\lambda _{b}] &=&\widehat{F}_{ab}-K_{ab} \\
&=&\widehat{F}_{ab}+\Phi _{ab}
\end{eqnarray*}
and therefore computes the field strength with the addition of the correct
central term. Moreover, for any function $f$ which transforms in the adjoint
representation of the non--commutative gauge group, the commutator 
\begin{eqnarray*}
-i[\lambda _{a},f] &=&\partial _{a}f-i[\widehat{A}_{a},f] \\
&=&\widehat{D}_{a}f
\end{eqnarray*}
computes the covariant derivative $\widehat{D}_{a}f$. Therefore, any
expression involving products of covariant derivatives of the field
strength, with the addition of the central term $\Phi =-K$, can be expressed
in terms of $\star $ products of the functions $\lambda _{a}$ -- for
example, an expression like $(\widehat{F}+\Phi )_{ab}\star \widehat{D}_{c}(%
\widehat{F}+\Phi )_{de}$ can be rewritten as $i[\lambda _{a},\lambda
_{b}]\star \lbrack \lambda _{c},[\lambda _{d},\lambda _{e}]]$. We conclude
that the general form of the effective action $S$ is 
\begin{equation*}
\frac{1}{G_{s}}\sum_{n\text{ even}}\int d^{D}x\det {}^{%
{\frac12}%
}\Gamma \,\ \left( \lambda _{a_{1}}\star \cdots \star \lambda
_{a_{n}}\right) \,\eta ^{a_{1}\cdots a_{n}}
\end{equation*}
where the coefficients $\eta ^{a_{1}\cdots a_{n}}$ are constructed from the
matrix $\Gamma ^{ab}$. We may further manipulate the above equation using (%
\ref{eq2-300}) and (\ref{eq2-400}) and raising and lowering indices with the
matrix $\theta ,K$. We can then write 
\begin{equation}
S=\frac{1}{g_{s}}\sum_{n\text{ even}}\int d^{D}x\det {}^{%
{\frac12}%
}K\,\ \left( \lambda ^{1}\star \cdots \star \lambda ^{n}\right) \,\eta
_{1\cdots n}  \label{eq3-100}
\end{equation}
We have used a compact notation for indices, which will be used extensively
in the sequel, substituting $a_{1}\rightarrow 1$, $a_{2}\rightarrow 2$, $%
\cdots $. Moreover, the symbol $\eta _{1\cdots n}$ represents the tensor
which is built from the matrix $\gamma _{ab}$ exactly as $\eta ^{1\cdots n}$
is constructed starting from $\Gamma ^{ab}$. For example, if $\eta
^{1234}=\Gamma ^{13}\Gamma ^{24}$ then $\eta _{1234}=\gamma _{13}\gamma
_{24} $.\medskip

\noindent {\small REMARK . Let us recall that, in the general }$U\left(
N\right) ${\small \ effective action, the distinction between derivative and
non--derivative terms is ambiguous, since a commutator of covariant
derivatives }$\left( D_{a}D_{b}-D_{b}D_{a}\right) \cdots ${\small \ is
equivalent to a commutator with the field strength }$\left[ F_{ab},\cdots %
\right] ${\small . In fact, when one writes the action in matrix form as in (%
\ref{eq3-100}), all terms (derivative and non--derivative) are included on
an equal footing. In particular, the ambiguity discussed above becomes
naturally the Jacobi identity of the commutator }$\left[ \lambda
^{a},\lambda ^{b}\right] ${\small .}

\subsection{\label{giiuact}Gauge invariance and invariance under addition of
a central term}

In the previous subsection we have rewritten the action $S$ in the compact
form (\ref{eq3-100}), which is more suited for discussing the action in its
entirety, including terms with arbitrary powers of the field strength and
with arbitrary number of derivatives. On the other hand, the gauge
invariance of the original action is not immediately transparent in this new
notation, and one needs to restate the requirement of gauge invariance in
terms of the tensors $\eta _{1\cdots n}$. This is easily done by noting that
the functions $\lambda _{a}$, when used in covariant expressions, always
appear within commutators, and therefore adding a constant $\varepsilon _{a}$
to the function $\lambda _{a}$ does not change the action. Gauge invariance
becomes then, within the matrix formulation (\ref{eq3-100}) of the action,
invariance under translations $\lambda ^{a}\rightarrow \lambda
^{a}+\varepsilon ^{a}$. This requirement quickly translates into the
following algebraic relation which must be satisfied by the tensors $\eta $%
\begin{equation}
\eta _{123\cdots n}+\eta _{213\cdots n}+\eta _{231\cdots n}+\cdots +\eta
_{23\cdots 1n}+\eta _{23\cdots n1}=0.  \label{eq1000}
\end{equation}
We will call tensors satisfying the above equation \textit{gauge invariant}
(GI).

In order to rewrite the effective action as a matrix action, we had fixed,
in section \ref{ama}, the central term $\Phi $ to $-K$. We must then require
by hand that the action (\ref{eq3-100}) be independent of the choice of
central term. This requirement will again be written as an algebraic
identity involving the tensors $\eta _{1\cdots n}$.

Let us then add a small central term $\kappa _{ab}\,$ to $\widehat{F}_{ab}$,
and at the same time subtract the same $\kappa _{ab}$ from the effective
metric $\Gamma _{ab}$. The two effects must compensate each other, yielding
a vanishing total variation of the action. Adding $\kappa $ to $\widehat{F}$
means that 
\begin{equation}
-i[\lambda ^{a},\lambda ^{b}]\rightarrow -i[\lambda ^{a},\lambda
^{b}]+\Delta ^{ab},  \label{eq4-100}
\end{equation}
with 
\begin{equation*}
\Delta =-\theta \kappa \theta .
\end{equation*}
Therefore, a term with $n+2$ coordinate functions $\lambda ^{a}$ will go
into a term with $n$ functions $\lambda ^{a}$. More precisely, the variation
of a term 
\begin{equation*}
\left( \lambda ^{1}\star \cdots \star \lambda ^{n+2}\right) \,\eta _{1\cdots
n+2}
\end{equation*}
will be of the form 
\begin{equation*}
-\left( \lambda ^{1}\star \cdots \star \lambda ^{n}\right) \,\left( 
\overline{\Delta }\eta \right) _{1\cdots n}
\end{equation*}
where $\left( \overline{\Delta }\eta \right) _{1\cdots n}$ \ is again gauge
invariant and depends on $\,\eta _{1\cdots n+2}$ and on $\Delta ^{ab}$. We
will show in the appendix (Lemma \ref{L-dbe}) that 
\begin{equation}
\left( \overline{\Delta }\eta \right) _{1\cdots n}=-\frac{i}{2}\Delta
^{ab}\left( \eta _{1\cdots nab}+\eta _{1\cdots anb}+\cdots \right) ,
\label{eq2000}
\end{equation}
where $\cdots $ indicates all the terms with the indices $1,\cdots ,n$ in
increasing order, and the two contracted indices $a,b$ in all possible
positions with $a$ preceding $b$. Let us just note that, for $n=2$, the
above result follows from (\ref{eq4-100}), since a gauge invariant $\eta
_{ab}$ is necessarily antisymmetric, and therefore $\lambda ^{a}\star
\lambda ^{b}\eta _{ab}=\frac{1}{2}[\lambda ^{a},\lambda ^{b}]\eta _{ab}$.

As noted previously, the variation (\ref{eq4-100}) must be compensated by a
corresponding change in the metric $\Gamma \rightarrow \Gamma -\kappa $. In
particular, in expression (\ref{eq3-100}) this will affect both the measure
of integration and the tensors $\eta _{1\cdots n}$. The measure changes by 
\begin{equation*}
\det {}^{%
{\frac12}%
}\Gamma \rightarrow \det {}^{%
{\frac12}%
}\Gamma \left( 1+\frac{1}{2}\kappa _{ab}\Gamma ^{ab}\right) ,
\end{equation*}
and the tensors by 
\begin{equation*}
\eta \rightarrow \eta -\kappa _{ab}\frac{\partial }{\partial \Gamma _{ab}}%
\eta .
\end{equation*}
Noting that $\kappa _{ab}\Gamma ^{ab}=\gamma _{ab}\Delta ^{ab}$ and that 
\begin{equation*}
-\kappa _{ab}\frac{\partial }{\partial \Gamma _{ab}}=(\gamma \Delta \gamma
)_{ab}\frac{\partial }{\partial \gamma _{ab}}
\end{equation*}
we then conclude that the variation of a term $\left( \lambda ^{1}\star
\cdots \star \lambda ^{n}\right) \eta _{1\cdots n}$ will be 
\begin{equation*}
\left( \lambda ^{1}\star \cdots \star \lambda ^{n}\right) \left( \overline{%
\delta }\eta \right) _{1\cdots n}
\end{equation*}
where 
\begin{equation*}
\overline{\delta }=\frac{1}{2}\left( \gamma _{ab}\Delta ^{ab}\right)
+(\gamma \Delta \gamma )_{ab}\frac{\partial }{\partial \gamma _{ab}}
\end{equation*}

Let us now combine the two variations. To this end, recall that the sum of
terms in (\ref{eq3-100}) runs only over even values of $n$. In particular we
will say that a tensor $\eta _{1,\cdots ,2L}$ with $2L$ indices is of level $%
L$, and we will denote it with $\eta ^{L}$. The operator $\overline{\Delta }$
lowers level by one, whereas $\overline{\delta }$ leaves the level
invariant. In order to balance the two variations $\overline{\Delta }$ and $%
\overline{\delta }$ and to have invariance under a change of the central
term we must then have that\footnote{%
We have checked invariance of the action under infinitesimal changes of $%
\Phi $ around $\Phi =-K$. On the other hand, this is sufficient, since
invariance under variation of the central term is a property of the \textit{%
structure }of the action, property which is independent of the specific
value of $\Phi $.} 
\begin{equation*}
\overline{\Delta }\eta ^{L+1}=\overline{\delta }\eta ^{L}.
\end{equation*}

\subsection{\label{swmfjs}The Seiberg--Witten map following Jurco and Schupp}

In section \ref{ama}, we have written the action $S$ in terms of the
functions $\lambda ^{a}$, which implicitly depend on the non--commutative
gauge potential $\widehat{A}=\widehat{A}_{SW}(A,\theta )$. In order to
analyze the independence of the action from the non--commutativity parameter 
$\theta $, it is convenient, as will become clear later, to rewrite $S$ in
terms of the $\theta $--independent abelian potential $A$. To this end, we
follow the analysis of Jurco and Schupp \cite{JS}, whose work describes the
Seiberg--Witten map $\widehat{A}_{SW}$ in an invariant way, which is best
suited for our purposes. Most of this section is then nothing but a review
of the ideas of \cite{JS}, rewritten in the notation of this paper.

First recall that, on the manifold $M$, one can define, in a natural way,
two distinct symplectic structures, defined by the two form $K$ and by the
combination 
\begin{equation*}
\omega =K+F,
\end{equation*}
where $F=dA$ is the usual abelian field strength. Since $F$ is exact, the
forms $K$ and $\omega $ define the same class in cohomology, and therefore,
by Darboux's lemma, there is a diffeomorphism $\lambda :M\rightarrow M$ such
that 
\begin{equation}
\lambda ^{\ast }\omega =K.  \label{eq5-500}
\end{equation}

Starting from the two symplectic structures $K$, $\omega $, one can, first
of all, define the corresponding Poisson brackets 
\begin{equation*}
\{f,g\}_{K}=\theta ^{ab}\partial _{a}f\partial _{b}g\,\ \ \ \ \ \ \ \ \ \ \
\ \ \{f,g\}_{\omega }=(\omega ^{-1})^{ab}\partial _{a}f\partial _{b}g.
\end{equation*}
It is clear that the two brackets are related by the diffeomorphism $\lambda 
$. More precisely, for any two functions $f,g$, one has the trivial identity 
\begin{equation}
\lambda ^{\ast }\{f,g\}_{\alpha }=\{\lambda ^{\ast }f,\lambda ^{\ast
}g\}_{\theta }  \label{eq5-100}
\end{equation}
where $\lambda ^{\ast }f=f\circ \lambda $.

From the two symplectic structures $K$ and $\omega $ one can also construct,
following Kontsevich \cite{K}, associated star--products 
\begin{equation*}
\star _{K}\,\ \ \ \ \ \ \ \ \ \ \ \ \ \ \ \ \ \ \ \star _{\omega }
\end{equation*}
In particular, $\star _{K}$ is nothing but the usual Moyal product, since in
the coordinates $x^{a}$ the symplectic structure $K$ is constant. The
product $\star _{\omega }$ is, on the other hand, the full product of
Kontsevich, which is expressed in terms of a complicated diagrammatic
expression involving derivatives of the Poisson structure $\omega ^{-1}$,
and for which there is an elegant path--integral expression, by Cattaneo and
Felder \cite{CF}. In what follows, we will not need the explicit form for $%
\star _{\omega }$. On the other hand, since $K$ and $\omega $ are related by
diffeomorphism, it is a general result of Kontsevich that the two products $%
\star _{K}$ and $\star _{\omega }$ are equivalent. More precisely, there is
a map $T$ defined on functions such that, for and functions $f$ and $g$, 
\begin{equation}
T\left( f\star _{\omega }g\right) =Tf\star _{K}Tg.  \label{eq5-200}
\end{equation}
The above expression is the analogue of expression (\ref{eq5-100}), and in
fact one can show that\footnote{%
We thank A. Cattaneo for pointing out that (\ref{eq5-900}) is a simple
consequence of formality, as defined in \cite{K}.} 
\begin{equation}
T=\lambda ^{\ast }(1+\cdots ),  \label{eq5-900}
\end{equation}
where $\cdots $ are higher order terms (more precisely, if we replace $%
K,\omega \rightarrow \frac{1}{\hbar }K,\frac{1}{\hbar }\omega $, then the
terms in $\cdots $ are higher order in $\hbar $).

Given these facts, one can define, in terms of $T$, the Seiberg--Witten map
as follows 
\begin{equation*}
\lambda ^{a}=Tx^{a}=x^{a}+\theta ^{ab}\widehat{A}_{b}.
\end{equation*}
We need to check that the map $\widehat{A}_{SW}$ implicitly defined above
maps gauge orbits of the abelian theory to gauge orbits of the
non--commutative theory. On one side, it is clear that different abelian
potentials $A$ which are gauge--equivalent do give the same map $\widehat{A}$%
, since $T$ is only defined in terms of the combination $\omega $, which is
itself gauge--invariant. Moreover, on the non--commutative side, we note
that the map $T$ defined in (\ref{eq5-200}) is only defined up to
transformations 
\begin{equation}
Tf\rightarrow \Lambda \star _{K}Tf\star _{K}\Lambda ^{-1},  \label{eq5-400}
\end{equation}
which leave (\ref{eq5-200}) invariant\footnote{%
The infinitesimal version of equation (\ref{eq5-400}) is given by $%
Tf\rightarrow Tf+[\rho ,Tf]_{K}$. This change of $T$ is analogous to the
fact that the map $\lambda $ is defined up to symplectomorphisms of the
manifold $(M,K)$. In fact, if $\chi :M\rightarrow M$ is such that $\chi
^{\ast }K=K$, then the composite map $\lambda \circ \chi $ still satisfies (%
\ref{eq5-500}). Recalling that symplectomorphisms are generated by
Hamiltonian flows, the change $\lambda ^{\ast }\rightarrow \left( \lambda
\circ \chi \right) ^{\ast }$ is given infinitesimally by $\lambda ^{\ast
}f\rightarrow \lambda ^{\ast }f+\{\rho ,\lambda ^{\ast }f\}_{K}$.}. This in
turn generates gauge transformations 
\begin{equation*}
\widehat{A}_{a}\rightarrow \widehat{A}_{a}+i\Lambda \star _{K}\partial
_{a}\Lambda ^{-1}+\Lambda \star _{K}\widehat{A}_{a}\star _{K}\Lambda ^{-1},
\end{equation*}
therefore showing that the transformation $\widehat{A}_{SW}$ does map gauge
orbits into gauge orbits.

\subsection{Integration}

In the previous section we have reviewed the Jurco--Schupp construction of
the Seiberg--Witten map. In this section we wish to discuss some issues
about integration of functions over $M$ which are closely related to the
discussion in the previous section, and which will be important in our
subsequent discussion.

Corresponding to the two symplectic structures $K$, $\omega $, one has two
volume--forms on $M$, respectively $d^{D}x\,\det {}^{%
{\frac12}%
}K\,$\ and $d^{D}x\,\det {}^{%
{\frac12}%
}\omega $, which are related by the map $\lambda $. More specifically, if $f$
is a generic function which vanishes at infinity, using the fact that $%
\lambda ^{\ast }\omega =K$, it is immediate to show that 
\begin{equation*}
\int d^{D}x\,\det {}^{%
{\frac12}%
}K\,\ \lambda ^{\ast }f=\int d^{D}x\,\det {}^{%
{\frac12}%
}\omega \,\ f.
\end{equation*}
Similarly, we may consider, recalling from (\ref{eq5-900}) that $T\sim
\lambda ^{\ast }$, the corresponding integral of $Tf$. We then have, in
general, that 
\begin{equation}
\int d^{D}x\,\det {}^{%
{\frac12}%
}K\,\ Tf=\int d^{D}x\,V(\omega )\ f,  \label{eq6-100}
\end{equation}
where $V(\omega )$ is a volume element depending on $\omega $ and its
derivatives, of the general form 
\begin{equation*}
V(\omega )=\det {}^{%
{\frac12}%
}\omega \left( 1+\cdots \right) ,
\end{equation*}
where $\cdots $ denotes, as in (\ref{eq5-900}), higher order derivative
corrections in $\omega ^{-1}$ which vanish if $\omega $ is constant. Let me
note that, since $\int d^{D}x\,\det {}^{%
{\frac12}%
}K\,\ f\star _{K}g=\int d^{D}x\,\det {}^{%
{\frac12}%
}K\,\ g\star _{K}f$, the ambiguity (\ref{eq5-400}) in the definition of $T$
does not affect the definition (\ref{eq6-100}) of $V(\omega )$, which really
only depends on the symplectic structure $\omega $. Moreover, from the
definition (\ref{eq5-200}) of $T$, we have in general that 
\begin{equation*}
\int d^{D}x\,V(\omega )\,\ f\star _{\omega }g=\int d^{D}x\,V(\omega )\,\
g\star _{\omega }f.
\end{equation*}

The explicit form of $V(\omega )$ is not know. On the other hand we will
show in the rest of the paper that some properties of $V(\omega )$ can be
proven indirectly, and this will suffice for our purposes.

\subsection{\label{bma}Back to the matrix action}

In this section we use the results just discussed on the Seiberg--Witten map
and on integration to rewrite the action (\ref{eq3-100}) 
\begin{equation*}
\frac{1}{g_{s}}\sum_{n\text{ even}}\int d^{D}x\det {}^{%
{\frac12}%
}K\,\ \left( \lambda ^{1}\star _{K}\cdots \star _{K}\lambda ^{n}\right)
\,\eta _{1\cdots n}
\end{equation*}
in an almost final form (we are now showing in the star--products $\star
_{K} $ the explicit dependence on the symplectic structure). Using the facts
that $\lambda ^{a}=Tx^{a}$ and that $T\left( f\star _{\omega }g\right)
=Tf\star _{K}Tg,$ one quickly sees that 
\begin{equation*}
\left( \lambda ^{1}\star _{K}\cdots \star _{K}\lambda ^{n}\right) =T\left(
x^{1}\star _{\omega }\cdots \star _{\omega }x^{n}\right) .
\end{equation*}
Using then equation (\ref{eq6-100}) on integration one concludes that the
action (\ref{eq3-100}) can be rewritten as 
\begin{equation*}
S=\frac{1}{g_{s}}\sum_{n\text{ even}}\int d^{D}x\ V(\omega )\,\ \left(
x^{1}\star _{\omega }\cdots \star _{\omega }x^{n}\right) \,\eta _{1\cdots n}.
\end{equation*}
The above action is written \textit{almost} exclusively in terms of the
closed string parameters $A,g+B$ and $g_{s}$. Moreover it is explicitly a
gauge invariant function of $A$, since the dependence on the gauge potential
is uniquely through the gauge invariant expression $\omega =K+F$. On the
other hand, the action $S$ above still depends on the parameter $\theta $,
through the definition of $\omega $ -- which effects $\star _{\omega }$ and $%
V(\omega )$ -- and through the effective metric $\gamma =g+B-K$ \ -- which
is the building block for the tensors $\eta $. However the action $S$ must
be independent of $\theta $, and the analysis of this requirement will be
the subject of the rest of the paper.\bigskip

\noindent R{\small EMARK . We are now in a position to give some very
intuitive arguments for the appearance of the full Kontsevich product in the
effective action. The arguments which follow are vague and not precise. On
the other hand, they provide a useful intuition, which, if made rigorous,
could be of importance. }

{\small Let us start by recalling \cite{Trev,TA} that the effective action
can be considered as the partition function for an open string sigma model} 
\begin{equation}
S(A)\propto \int DX\,\ e^{-I_{S}-I_{A}}  \label{r2}
\end{equation}
{\small where} 
\begin{eqnarray*}
I_{S} &=&\frac{1}{\alpha ^{\prime }}\int_{\Sigma }g_{ab}\,\partial X^{a}%
\overline{\partial }X^{b} \\
I_{A} &=&\int_{\Sigma }B+\int_{\partial \Sigma }A
\end{eqnarray*}
{\small If we consider the naive limit }$\alpha ^{\prime }\rightarrow 0$%
{\small , the term }$I_{S}${\small \ dominates, and we should consider }$%
I_{A}${\small \ as a perturbation. On the other hand, we recall that, in 
\cite{SW}, Seiberg and Witten consider the limit }$\alpha ^{\prime
},g_{ab}\rightarrow 0${\small , with }$g_{ab}/\alpha ^{\prime }\rightarrow 0$%
{\small . Therefore, in this case, the dominating term is }$I_{A}${\small .
We also note that } 
\begin{equation}
I_{A}=\int_{\Sigma }\left( B+F\right)  \label{r1}
\end{equation}
{\small and that the above is nothing but the Cattaneo--Felder model \cite
{CF}} 
\begin{equation*}
\int \eta _{a}\wedge dX^{a}+\frac{1}{2}\alpha ^{ab}(X)\eta _{a}\wedge \eta
_{b}
\end{equation*}
{\small in the special case of invertible Poisson structure }$\alpha ^{ab}$%
{\small , with }$\alpha ^{-1}=B+F${\small . In this case, one can integrate
out the one--forms }$\eta _{i}${\small , which appear quadratically, and
recover (\ref{r1}). We recall that the perturbation theory of the
Cattaneo--Felder model generates the Kontsevich graphs, which are the basis
of the product }$\star _{B+F}${\small . One then expects (in an undoubtedly
vague way) to obtain effective actions based on the full Kontsevich product.
Moreover one expects to obtain, among the various products considered in 
\cite{K}, the simplest one, defined using the \textit{harmonic angle map}.
Along the same lines one could also expand (\ref{r2}) in powers of }$I_{S}$ 
{\small and obtain an expansion like} 
\begin{equation*}
\int DX\,\ e^{-I_{A}}I_{S}I_{S}\sim \int d^{D}x\ V(\omega )\,\
[x^{a},x^{b}]_{\omega }\star _{\omega }[x^{c},x^{d}]_{\omega }\,g_{ac}g_{bd}
\end{equation*}
{\small where the RHS above is nothing but the }$\widehat{F}^{2}${\small \
term in the action, which dominates in the }$\alpha ^{\prime }\rightarrow 0$ 
{\small limit of \cite{SW}}.

\subsection{\label{bict}Behavior at infinity and cyclic tensors}

We have seen that the general action describing the dynamics of gauge fields
is of the form 
\begin{equation}
\frac{1}{g_{s}}\sum_{n\text{ even}}\int d^{D}x\ V(\omega )\,\ \left(
x^{1}\star _{\omega }\cdots \star _{\omega }x^{n}\right) \,\eta _{1\cdots n}
\label{eq100}
\end{equation}
Let us now proceed by first concentrating on a single term in the sum (\ref
{eq100}). In particular let us focus on the expression 
\begin{equation*}
\eta (x)=x^{1}\star _{\omega }\cdots \star _{\omega }x^{n}\;\eta _{1\cdots n}
\end{equation*}
The above clearly defined a function $\eta $ of the coordinates $x$, which
is written in terms of $\omega ^{-1}$ and its derivatives. We will assume
throughout the paper that 
\begin{equation*}
F(x)\rightarrow 0\,\ \ \ \ \ \ \ \ \ \ \ \mathrm{when}\text{\ }x\rightarrow
\infty
\end{equation*}
This implies that, for large $x$, the symplectic structure $\omega
\rightarrow K$ becomes constant, and that the star--product $\star _{\omega
} $ becomes the Moyal product $\star _{K}$ with respect to $\theta $. In
general then, for $x\rightarrow \infty $, the function $\eta (x)$ is a
polynomial of degree $n$ in the coordinates $x^{a}$. If we assume further
that $\eta _{1\cdots n}$ is a gauge invariant tensor, then we can quickly
see that, again for $x\rightarrow \infty $, 
\begin{equation*}
\eta (x+\varepsilon )-\eta \left( x\right) =\varepsilon ^{1}x^{2}\star
_{K}\cdots \star _{K}x^{n}\;(\eta _{12\cdots n}+\eta _{21\cdots n}+\cdots )=0
\end{equation*}
and therefore the function $\eta (x)$ approaches a constant $\eta _{\infty }$
at infinity. Similarly, the volume form $V(\omega )$ converges to the
constant $\det {}^{%
{\frac12}%
}K$ as $x\rightarrow \infty $. It is then clear that the integral $\int
d^{D}x\,V(\omega )\,\eta $ in general diverges unless $\eta _{\infty }=0$.
In order to define the action properly we should replace 
\begin{equation}
\int d^{D}xV(\omega )\eta \rightarrow \int d^{D}xV(\omega )\left[ \eta -\eta
_{\infty }\right]  \label{eq500}
\end{equation}
thereby eliminating the infinities coming from integration over an infinite
world--volume. Let me note that, since $T1=1$, one has $\int d^{D}xV(\omega
)=\int d^{D}x\det {}^{%
{\frac12}%
}K$, and therefore the subtraction (\ref{eq500}) is independent of $F$.
Replacement (\ref{eq500}) is then nothing but a constant addition to the
action.

We will now show how the subtraction (\ref{eq500}) can be achieved in an
invariant way, without explicitly considering the behavior at infinity.
First let us recall that, for functions $f$ and $g$ which vanish at
infinity, we have that 
\begin{equation*}
\int d^{D}x\,V(\omega )\,\ f\star _{\omega }g=\int d^{D}x\,V(\omega )\,\
g\star _{\omega }f.
\end{equation*}
We are therefore tempted to say that the integral $\int d^{D}x\ V(\omega
)\,\ \left( x^{1}\star _{\omega }\cdots \star _{\omega }x^{n}\right) \,$ is
invariant under cyclic permutations of the indices $1,2,\cdots ,n$. This is,
on the other hand, not quite correct, since the coordinate functions $x^{a}$
which enter in expression (\ref{eq100}) clearly do not vanish for $%
x\rightarrow \infty $. Nonetheless let us, for the moment, blindly assume
cyclicity of the integral. We may then substitute, in expression (\ref{eq100}%
) for the action, the tensors $\eta _{1\cdots n}$ with the cyclically
symmetrized tensors 
\begin{equation}
\tau _{1\cdots n}=\frac{1}{n}\left( \eta _{1\cdots n}+\mathrm{cyc}_{1\cdots
n}\right) ,  \label{eq007}
\end{equation}
where $\mathrm{cyc}_{1\cdots n}$ denotes the sum over cyclic permutations of
the indices $1,\cdots ,n$. Gauge invariance of the tensors $\eta $ then
translates into the following algebraic property satisfied by the tensors $%
\tau $%
\begin{equation}
\tau _{123\cdots n}+\tau _{213\cdots n}+\tau _{231\cdots n}+\cdots +\tau
_{23\cdots 1n}=0.  \label{eq400}
\end{equation}
Note that the above expression is very similar to (\ref{eq1000}), with the
only difference that the \textit{moving }index $1$ runs only over the
cyclically independent orderings, and therefore the last term in (\ref
{eq1000}) is absent in (\ref{eq400}). Tensors which satisfy the above
relation will be called \textit{cyclic gauge invariant }(CGI). We leave the
proof of (\ref{eq400}) to the appendix (Lemma \ref{L-cgi}).

\ We may now consider, similarly to the previous analysis, the function 
\begin{equation*}
\tau (x)=x^{1}\star _{\omega }\cdots \star _{\omega }x^{n}\;\tau _{1\cdots n}
\end{equation*}
and in particular its behavior at infinity. As shown in the appendix (Lemma 
\ref{L-1}), for $n$ even (which is the case relevant to equation (\ref{eq100}%
)) one has that 
\begin{equation*}
\tau (x)\rightarrow 0
\end{equation*}
for $x\rightarrow \infty $. Therefore the integral 
\begin{equation*}
\int d^{D}xV(\omega )\tau
\end{equation*}
is well defined. Moreover we will show in the appendix (Lemma \ref{L-2})
that, generically, one has that 
\begin{equation*}
\int d^{D}xV(\omega )\tau =\int d^{D}xV(\omega )\left[ \eta -\eta _{\infty }%
\right]
\end{equation*}
so that we have lost nothing by assuming cyclicity\footnote{%
Let me note that, although the functions $\tau $ and $\eta -\eta _{\infty }$
have the same integral, and therefore define the same functional of $A$, one
has in general that $\tau \neq \eta -\eta _{\infty }$.}. In fact, using
cyclic gauge invariant tensors, the subtraction which was needed in (\ref
{eq500}) in order to properly define the action $S(A)$ is automatically
incorporated into the formalism. We will therefore consider, from now on,
the final form of the action 
\begin{equation}
S=\frac{1}{g_{s}}\sum_{n\text{ even}}\int d^{D}x\ V(\omega )\,\ \left(
x^{1}\star _{\omega }\cdots \star _{\omega }x^{n}\right) \tau _{1\cdots n}
\label{eq600}
\end{equation}
where the tensors $\tau $ are cyclic gauge invariant.

We have seen in section \ref{giiuact} that, in order for the tensors $\eta $
to define an action, they had to be gauge invariant and they had to satisfy $%
\overline{\Delta }\eta ^{L+1}=\overline{\delta }\eta ^{L}$. These two
properties impose restrictions on the cyclically symmetrized tensors $\tau $%
. Gauge invariance of the $\eta $'s implies cyclic gauge invariance of the $%
\tau $'s. The equation $\overline{\Delta }\eta ^{L+1}=\overline{\delta }\eta
^{L}$ implies a similar equation for the $\tau $'s, which we now describe.

Consider a gauge invariant tensor $\eta _{1\cdots n+2}$. We can construct,
given $\Delta ^{ab}$ and equation (\ref{eq2000}), the tensor $\left( 
\overline{\Delta }\eta \right) _{1\cdots n}$, which is also gauge invariant
for any choice of $\Delta ^{ab}$. We may then consider the cyclically
symmetric combination $g_{1\cdots n}=\frac{1}{n}\left( \overline{\Delta }%
\eta _{1\cdots n}+\mathrm{cyc}_{1\cdots n}\right) $, which will be a cyclic
gauge invariant tensor. At first sight the tensor $g_{1\cdots n}$ is a
function of the original tensor $\eta _{1\cdots n+2}$, but, as we will prove
in the appendix (Lemma \ref{L-3}), it is actually just a function of the
cyclically symmetrized tensor $\tau _{1\cdots n+2}$. We will then denote the
tensor $g_{1\cdots n}$ with $\left( \overline{\Delta }\tau \right) _{1\cdots
n}$, where 
\begin{equation*}
\left( \overline{\Delta }\tau \right) _{1\cdots n}=-\frac{i}{2}\left( \frac{%
n+2}{n^{2}}\right) \Delta ^{ab}\left[ n\tau _{1\cdots nab}+\left( n-1\right)
\tau _{1\cdots anb}+\cdots +0\cdot \tau _{a1\cdots nb}\right] +\mathrm{cyc}%
_{1\cdots n}
\end{equation*}
Nothing on the other hand needs to be altered in the definition of the
operator 
\begin{equation*}
\overline{\delta }=\frac{1}{2}\left( \gamma _{ab}\Delta ^{ab}\right)
+(\gamma \Delta \gamma )_{ab}\frac{\partial }{\partial \gamma _{ab}}
\end{equation*}
which commutes with the symmetrization (\ref{eq007}). We then have the
requirement on the tensors $\tau $%
\begin{equation}
\overline{\Delta }\tau ^{L+1}=\overline{\delta }\tau ^{L}.  \label{lula}
\end{equation}
\bigskip

\noindent E{\small XAMPLE. Let us compute, as an important example, the
first non--vanishing tensor }$\tau ^{2}${\small . If we consider the
expansion of the Born--Infeld action (with }$\Omega =\widehat{F}-K${\small )}
\begin{equation*}
\sqrt{\det (\Gamma +\Omega )}\propto 1-\frac{1}{2}\Gamma ^{ab}\Omega _{ab}+%
\frac{1}{4}\Gamma ^{ac}\Gamma ^{db}\Omega _{ab}\Omega _{cd}+\frac{1}{8}%
\Gamma ^{ab}\Omega _{ab}\Gamma ^{cd}\Omega _{cd}
\end{equation*}
{\small we can quickly see that} 
\begin{eqnarray*}
\eta _{12} &=&\frac{i}{2}\left( \gamma _{12}-\gamma _{21}\right) \\
\eta _{1234} &=&-\frac{1}{4}\left( \gamma _{13}\gamma _{42}-\gamma
_{23}\gamma _{41}+\gamma _{31}\gamma _{24}-\gamma _{32}\gamma _{14}\right) \\
&&-\frac{1}{8}\left( \gamma _{12}\gamma _{34}-\gamma _{21}\gamma
_{34}+\gamma _{12}\gamma _{43}-\gamma _{21}\gamma _{43}\right) .
\end{eqnarray*}
{\small We may then compute the cyclically symmetrized tensors }$\tau $%
{\small . Clearly }$\tau _{12}=0${\small . A simple computation also shows
that } 
\begin{equation}
\tau _{1234}=\frac{1}{4}g_{12}g_{34}+\frac{1}{4}g_{14}g_{23}-\frac{1}{2}%
g_{13}g_{24},  \label{lala}
\end{equation}
{\small where, we recall,} 
\begin{equation*}
g_{ab}=\frac{1}{2}\left( \gamma _{ab}+\gamma _{ba}\right)
\end{equation*}
{\small is the symmetric part of the tensor }$\gamma _{ab}${\small . One can
also check, given (\ref{lala}), that} 
\begin{equation*}
\overline{\Delta }\tau ^{2}=0,
\end{equation*}
{\small which is consistent with (\ref{lula}) and the fact that }$\tau
^{1}=0 ${\small . }

\section{\label{S4}Operator Description}

In this section we leave momentarily the analysis of the effective action $S$%
, and we develop some formal tools which will allow us both to rewrite the
various equations in a more compact and natural way, and also to tackle the
problem of the independence of the action $S$ on the parameter $\theta $.

First we analyze, within a general framework, the description of the action
using operators. We then study a simple $2$--dimensional example, which is
not directly relevant to our more general situation, but which is completely
tractable and which will be a useful frame of reference in discussing the
general case. We then move on to the situation most relevant for this paper,
and discuss, in that case, the generalization of the results obtained in the 
$2$--dimensional setting.

\subsection{\label{orsp}Operator representation of star products}

Let us consider first a flat symplectic structure $K$. We introduce a set of
operators $J^{a}$ with commutation relations $[J^{a},J^{b}]=i\theta ^{ab}$,
which can be represented on the Hilbert space $\mathcal{H}=L^{2}(\mathbb{R}%
^{D/2})$ as linear combinations of the standard $p,q\,$\ operators in
quantum mechanics (recall that we are assuming $\theta $ invertible). To any
function $f$ on phase space $M=\mathbb{R}^{D}$ we can associate, using Weyl
ordering, an operator $Q_{K}(f)$ acting on $\mathcal{H}$. It is then well
known that, if $f$ and $g$ are two generic functions, then $%
Q_{K}(f)Q_{K}(g)=Q_{K}(f\star _{K}g)$. Moreover, if $f$ vanishes at
infinity, one also has (up to an overall constant $\left( 2\pi \right)
^{D/2} $ which can be, for example, reabsorbed in the definition of the
trace) that $\mathrm{Tr}(Q_{K}(f))=\int d^{D}x$ $\det {}^{%
{\frac12}%
}K\,f$. We will call $Q$ a quantization map.

Consider now a general symplectic structure $\omega $. Since any two
symplectic structures on $M$ are related by a diffeomorphism, one can follow
section \ref{swmfjs} and find a map $T$ on the space of functions such that,
for general $f,g$, one has $T\left( f\star _{\omega }g\right) =Tf\star
_{K}Tg $. One may then define a new quantization map $Q_{\omega }$, related
to the symplectic structure $\omega $, by the following relation 
\begin{equation}
Q_{\omega }(f)=Q_{K}(Tf).  \label{eq10-300}
\end{equation}
It is then simple to show that 
\begin{equation}
Q_{\omega }(f)Q_{\omega }(g)=Q_{\omega }(f\star _{\omega }g)  \label{eq3-200}
\end{equation}
and that 
\begin{equation}
\mathrm{Tr}(Q_{\omega }f))=\int d^{D}x\text{ }V(\omega )\,\,f.
\label{eq3-400}
\end{equation}

Let us note that, for any fixed symplectic form $\omega $, the map $%
Q_{\omega }$ is actually only defined up to conjugation. Recall first that
the map $T$ is defined up to a redefinition of the form $Tf\rightarrow 
\widetilde{T}f=\Lambda \star _{K}Tf\star _{K}\Lambda ^{-1}$. Using $%
\widetilde{T}$ in equation (\ref{eq10-300}), and letting $\chi
=T^{-1}\Lambda $, we obtain a new map $\widetilde{Q}_{\omega }$ which reads,
in terms of the original $Q_{\omega }$, 
\begin{eqnarray}
\widetilde{Q}_{\omega }(f) &=&Q_{\omega }(\chi )Q_{\omega }(f)Q_{\omega
}^{-1}(\chi )  \label{eq10-700} \\
&=&Q_{\omega }(\chi \star _{\omega }f\star _{\omega }\chi ^{-1}).  \notag
\end{eqnarray}

We now use this notation to rewrite the action $S$ in a compact and
invariant way. First define the operators 
\begin{equation*}
X^{a}=Q_{K}(\lambda ^{a})=Q_{\omega }(x^{a}).
\end{equation*}
Then the action (\ref{eq600}) can then be compactly written as 
\begin{equation*}
S=\frac{1}{g_{s}}\sum_{n\text{ even}}\mathrm{Tr}\left( X^{1}\cdots
X^{n}\right) \tau _{1\cdots n}.
\end{equation*}
The ambiguity (\ref{eq10-700}) is reflected in a possible redefinition $%
X^{a}\rightarrow OX^{a}O^{-1}$, which on the other hand does not affect the
action.

The action $S$ written above implicitly depends on a specific choice of
non--commutativity parameter $\theta $. The dependence is two--fold. On one
hand the tensors $\tau $ are built starting from the metric $\gamma $, which
linearly depends on $K$. On the other hand, the parameter $K$ enters into
the definition of the symplectic structure $\omega $, and therefore it
implicitly determines the operators $X^{a}=Q_{\omega }(x^{a})$. It is then
clear that we need to understand the variation of the quantization map $%
Q_{\omega }$, when we add to $\omega _{ab}$ a \textit{constant}
antisymmetric matrix $\Delta _{ab}$. This is the subject of the next two
sections. In particular, in the next section, we analyze this problem within
a simple two--dimensional model, related to the general framework which we
developed above. In this two--dimensional model the variation of $Q_{\omega
} $ for $\omega \rightarrow \omega +\Delta $ can be completely analyzed.
Moreover the solution will give us the correct ansatz to tackle the general
problem.

\subsection{\label{s2de}A simple $2$--dimensional example}

The general framework of this section follows closely \cite{C1}. We consider
the space $\mathcal{V}$ of complex functions on the complex plane $\mathbb{C}
$, and the subspace $\mathcal{H}\subset \mathcal{V}$ of holomorphic
functions. We then make $\mathcal{V}$ into a Hilbert space by choosing a
real positive function $C$ on the complex plane and by letting the inner
product of two function $\psi ,\phi \in \mathcal{V}$ be given by 
\begin{equation*}
\left\langle \psi |\phi \right\rangle =\int d^{2}z\,C\,\overline{\psi }\phi .
\end{equation*}
Associated to $C$ we have a natural symplectic form $i\omega \,dz\wedge d%
\overline{z}$ on $\mathbb{C}$ with 
\begin{equation*}
\omega =-\partial \overline{\partial }\ln C\text{.}
\end{equation*}
One may also consider the orthogonal projection 
\begin{equation*}
\pi :\mathcal{V}\rightarrow \mathcal{H},
\end{equation*}
which clearly depends on the choice of inner product on $\mathcal{V}$, and
therefore on $C$. Given a generic function $f$, we may then consider the
corresponding operator 
\begin{equation*}
Q_{\omega }(f):\mathcal{H}\rightarrow \mathcal{H}
\end{equation*}
defined by 
\begin{equation*}
Q_{\omega }(f)\eta =\pi \,f\,\eta .
\end{equation*}
for $\eta \in \mathcal{H}$. In words, the operator $Q_{\omega }(f)$ first
multiplies pointwise by $f$ \ -- which is not assumed to be holomorphic --
and then extracts the holomorphic part of the resulting function using the
projector $\pi $. It is shown in \cite{C1} that, given two functions $f,g$, 
\begin{eqnarray}
Q_{\omega }(f)Q_{\omega }(g) &=&Q_{\omega }(f\star _{\omega }g)
\label{eq10-100} \\
\mathrm{Tr}_{\mathcal{H}}(Q_{\omega }(f)) &=&\int d^{2}z\,V(\omega )\,f 
\notag
\end{eqnarray}
where $\star _{\omega }$ is a \textit{holomorphic} star product ($f\star
_{\omega }g=fg$ if either $\overline{f}$ or $g$ are holomorphic) related to $%
\omega $ and $V(\omega )=\omega (1+\cdots )$. The operator $Q_{\omega }$
actually depends on $C$, and not simply on $\omega $. On the other hand, a
change of $C$ which leaves $\omega $ invariant changes the operators $%
Q_{\omega }(f)$ by conjugation, exactly as in (\ref{eq10-700}).

We may now consider the variation $\omega \rightarrow \omega +\Delta $, with 
$\Delta $ an infinitesimal constant. This corresponds to 
\begin{equation*}
C\rightarrow \widetilde{C}=Ce^{-z\overline{z}\Delta }.
\end{equation*}
We need to understand the change in $\pi $. Let then $\left\langle
|\right\rangle $ denote the original inner product with $C$, and let $%
|n\rangle $ be an orthonormal basis for $\mathcal{H}$. Then $\pi
=\sum_{n}|n\rangle \langle n|$. It is easy to show that the vectors 
\begin{equation*}
|\widetilde{n}\rangle =|n\rangle +\frac{\Delta }{2}\sum_{m}|m\rangle
\left\langle m|z\overline{z}|n\right\rangle
\end{equation*}
satisfy $\left\langle \widetilde{n}|e^{-z\overline{z}\Delta }|\widetilde{m}%
\right\rangle =\delta _{n,m}$ (to first order in $\Delta $) and that the new
projection $\widetilde{\pi }$ related to $\widetilde{C}$ is given by $%
\widetilde{\pi }=\sum_{n}|\widetilde{n}\rangle \langle \widetilde{n}|e^{-z%
\overline{z}\Delta }$. Using the explicit expression for $|\widetilde{n}%
\rangle $, one can then show that 
\begin{equation*}
\widetilde{\pi }=\pi +\Delta \left( \pi z\overline{z}\pi -\pi z\overline{z}%
\right) .
\end{equation*}
Let now $f$ be a generic function and $F=Q_{\omega }(f)=\pi f$. Noting that $%
\pi z=z$ one can show that 
\begin{eqnarray*}
Q_{\omega +\Delta }(f) &=&\widetilde{\pi }f=F+\Delta (\pi \overline{z}z\pi
f-\pi \overline{z}fz) \\
&=&F+\Delta (\overline{Z}ZF-Q_{\omega }(\overline{z}f)Z)
\end{eqnarray*}
where $Z=Q_{\omega }(z)$ and $\overline{Z}=Q_{\omega }(\overline{z})$. Using
the fact that the star product is holomorphic and that $Q_{\omega }(%
\overline{z}f)=Q_{\omega }(\overline{z}\star _{\omega }f)=\overline{Z}F$, we
arrive at the result 
\begin{equation*}
Q_{\omega +\Delta }(f)=F+\Delta (\overline{Z}ZF-\overline{Z}FZ).
\end{equation*}

The exact form of the above equation depends on the specific quantization
model we chose to analyze. On the other hand the general lesson that should
be drawn is that the variation $Q_{\omega +\Delta }(f)-Q_{\omega }(f)$
contains $F$ and two powers of the coordinate operators $X^{a}=Q_{\omega
}(x^{a})$, with some ordering. We will use this intuition in the next
section to compute $Q_{\omega +\Delta }(f)$ in the setting of section \ref
{orsp}.

\subsection{\label{qmgc}The quantization map $Q_{\protect\omega +\Delta }$
in the general case}

We have seen, from the previous example, that, if $f$ is a generic function,
then the variation $Q_{\omega +\Delta }(f)-Q_{\omega }(f)$ is proportional
to $\Delta _{ab}$ and to the product of 
\begin{equation*}
F=Q_{\omega }(f)
\end{equation*}
and of two coordinate operators $X^{a}=Q_{\omega }(x^{a})$ in some specific
ordering. More precisely, the operator $Q_{\omega +\Delta }(f)$ must be
equal to $F+\frac{i}{4}\Delta _{ab}(aX^{a}X^{b}F+bFX^{a}X^{b}+cX^{a}FX^{b})$%
, for some choice of the coefficients $a,b,c$. It is shown in the appendix
(Lemma \ref{L-5}) that, in the case in which the underlying star--product is
that of Kontsevich, the correct coefficients are $a=b=1$, and $c=-2$.

We have then the basic relation 
\begin{equation}
Q_{\omega +\Delta }(f)=F+\frac{i}{4}\Delta _{ab}\left(
X^{a}X^{b}F+FX^{a}X^{b}-2X^{a}FX^{b}\right) .  \label{eq3-300}
\end{equation}
The above can be alternatively rewritten as 
\begin{equation}
Q_{\omega +\Delta }(f)=Q_{\omega }(f+Rf).  \label{eq3-10000}
\end{equation}
where 
\begin{equation}
Rf=\frac{i}{4}\Delta _{ab}\left( x^{a}\star _{\omega }x^{b}\star _{\omega
}f+f\star _{\omega }x^{a}\star _{\omega }x^{b}-2x^{a}\star _{\omega }f\star
_{\omega }x^{b}\right) .  \label{eqa100}
\end{equation}
As a consequence of the above facts, we have the following two results.
First of all combining (\ref{eq3-10000}) and (\ref{eq3-200}) we obtain 
\begin{equation}
f\star _{\omega +\Delta }g=f\star _{\omega }g-\frac{i}{2}\Delta _{ab}\left[
x^{a},f\right] _{\omega }\star _{\omega }\left[ x^{b},g\right] _{\omega }.
\label{r3}
\end{equation}
Also, taking the trace of (\ref{eq3-300}) and using (\ref{eq3-400}) we
deduce that, for functions $f$ vanishing at infinity, 
\begin{equation}
\int d^{D}x\text{ }V(\omega +\Delta )\,f=\int d^{D}x\,\ V(\omega )\,\left[
f+Rf\right] .  \label{eqa200}
\end{equation}
\bigskip

\noindent R{\small EMARK . The fact that the variation of the star--product }%
$\star _{\omega }${\small \ under the change }$\omega \rightarrow \omega
+\Delta ${\small \ is given by an expression involving a quadratic
combination (\ref{r3}) of the coordinate functions }$x^{a}${\small \ can
also be understood intuitively using the Cattaneo--Felder model. As in the
remark in section \ref{bma}, the argument is very vague, but it would be
very useful to make it rigorous.}

{\small The star product }$f\star _{\omega }g${\small \ is given by the disk
expectation value }$\left\langle f(X(0))g(X(1))\right\rangle ${\small , with
weight }$\int DX\,\ \exp \left( -\int_{\Sigma }\omega \right) ${\small .
Therefore, under the change }$\omega \rightarrow \omega +\Delta ${\small ,
one has that }$f\star _{\omega +\Delta }g-f\star _{\omega }g${\small \ is
given by }$-\int_{\Sigma }\left\langle \Delta \,f(X(0))g(X(1))\right\rangle $%
{\small . But }$\Delta =\frac{1}{2}\Delta _{ab}dX^{a}\wedge dX^{b}${\small ,
thus giving a quadratic expression in the coordinate functions.}

\section{\label{S5}Invariance of the Action}

\subsection{\label{iuctbc}Invariance under a change in $\protect\theta $ and
the basic commutator}

We now have all the tools that we need to tackle our main problem. Let me
first recall were we stand. The action is given by 
\begin{equation}
S=\frac{1}{g_{s}}\sum_{n\text{ even}}\mathrm{Tr}\left( X^{1}\cdots
X^{n}\right) \tau _{1\cdots n}.  \label{eq50-200}
\end{equation}
where $X^{a}=Q_{\omega }(x^{a})$ and the tensors $\tau $ are cyclic gauge
invariant tensors built from $\gamma $. The tensors $\tau $ satisfy the
consistency condition $\overline{\delta }\tau ^{L}=\overline{\Delta }\tau
^{L+1}$, where $\tau ^{L}$ denotes the tensor at level $L$, with $2L$
indices.

The action $S$ depends on a specific choice of non--commutativity parameter $%
\theta $. The dependence is two--fold, through $\gamma =g+B-K$ and through $%
\omega =K+F$. We have argued in previous sections that the total dependence
on $\theta $ should vanish, and therefore the two variations of the action
under a change of $K$ should compensate each other and sum to zero. This
clearly imposes additional restrictions on the possible forms of the tensors 
$\tau $, which we now analyze.

Let us start by considering an infinitesimal variation of $K$ given by 
\begin{equation*}
K\rightarrow K+\Delta ,
\end{equation*}
where $\Delta _{ab}$ is an arbitrary antisymmetric constant matrix. The
metric $\gamma $ then changes as follows 
\begin{equation*}
\gamma \rightarrow \gamma -\Delta
\end{equation*}
therefore implying a change in the tensors $\tau $ given by 
\begin{equation*}
\tau \rightarrow \tau -\delta \tau ,
\end{equation*}
where $\delta $ is the differential operator defined by 
\begin{equation*}
\delta =\Delta _{ab}\frac{\partial }{\partial \gamma _{ab}}.
\end{equation*}
On the other hand, the symplectic structure $\omega $ changes by a constant
term 
\begin{equation*}
\omega \rightarrow \omega +\Delta .
\end{equation*}
Then, as discussed in the previous section, the coordinate operators $X^{c}$
change as 
\begin{equation}
X^{c}\rightarrow X^{c}+\frac{i}{4}\Delta _{ab}\left(
X^{a}X^{b}X^{c}+X^{c}X^{a}X^{b}-2X^{a}X^{c}X^{b}\right) .  \label{eq40-100}
\end{equation}

We can now discuss the variation of the term $\mathrm{Tr}\left( X^{1}\cdots
X^{n-2}\right) \tau _{1\cdots n-2}$. It will consists of two parts, coming
from the variation of the tensor $\tau $ and from the variation of the
coordinate functions $X^{a}$. The first part is simply 
\begin{equation*}
-\mathrm{Tr}\left( X^{1}\cdots X^{n-2}\right) \left( \delta \tau \right)
_{1\cdots n-2}.
\end{equation*}
The second will involve the trace of $n$ coordinates, and will be of the
general form 
\begin{equation*}
\mathrm{Tr}\left( X^{1}\cdots X^{n}\right) \left( \Delta \tau \right)
_{1\cdots n},
\end{equation*}
where the tensor $\left( \Delta \tau \right) _{1\cdots n}$ is built from $%
\tau _{1\cdots n-2}$ and from $\Delta _{ab}$. Applying equation (\ref
{eq40-100}) to the expression $\mathrm{Tr}\left( X^{1}\cdots X^{n-2}\right) $
and rearranging cyclically under the trace it is easy to show that 
\begin{equation}
\left( \Delta \tau \right) _{1\cdots n}=\frac{i}{2}\left( \frac{n-2}{n}%
\right) \left( \Delta _{12}\tau _{345\cdots n}-\Delta _{13}\tau _{245\cdots
n}\right) +\mathrm{cyc}_{1\cdots n}  \label{eq50-100}
\end{equation}
First we note that the above tensor $\Delta \tau $ is cyclic gauge invariant
for any choice of $\Delta _{ab}$. This fact is proved in the appendix (Lemma 
\ref{L-6}). Moreover, the fact that both $\tau $ and $\Delta \tau $ are
cyclic gauge invariant is crucial in a more careful derivation of (\ref
{eq50-100}). In fact, the use of cyclic symmetry under the trace is formally
correct, and does give the correct answer. On the other hand, one needs to
check that the formal manipulations can be justified, since the coordinate
functions do not vanish at infinity, and therefore one might forget
important boundary terms. The detailed proof of equation (\ref{eq50-100}) is
again given in Lemma \ref{L-6}.

We can then finally state the main algebraic equation which must be
satisfied by the tensors $\tau $ in order for the action (\ref{eq50-200}) to
be invariant under changes of $\theta $. Again indicating with $L$ the level
of a tensor $\tau _{1,\cdots ,2L}$ with $2L$ indices, we have the basic
equation 
\begin{equation}
\Delta \tau ^{L}=\delta \tau ^{L+1}  \label{eqinv1}
\end{equation}
which must be considered together with the equation 
\begin{equation}
\overline{\Delta }\tau ^{L+1}=\overline{\delta }\tau ^{L}  \label{eqinv2}
\end{equation}
previously analyzed.

The above two equations involve the basic operators $\Delta ,\delta ,%
\overline{\Delta },\overline{\delta }$, which in general depend on two
distinct antisymmetric matrices $\Delta _{ab}$ and $\Delta ^{ab}$. Without
loss in generality we can assume that 
\begin{equation*}
\Delta _{ac}\Delta ^{cb}=\delta _{a}^{b}.
\end{equation*}
In this case one has simple relations between the operators $\delta ,%
\overline{\delta }$ and $\Delta ,\overline{\Delta }$, which we now describe.
First we introduce the operator $N$ which simply counts the number of
indices of a tensor, and which is defined by 
\begin{equation*}
N\tau _{1\cdots n}=n\cdot \tau _{1\cdots n}.
\end{equation*}
It is then easy to see that, since $\tau _{1\cdots 2L}$ is built from $L$
copies of $\gamma $, one has that 
\begin{equation*}
2\gamma _{ab}\frac{\partial }{\partial \gamma _{ab}}=N.
\end{equation*}
Using the above relation one then discovers quickly that 
\begin{equation*}
\lbrack \delta ,\overline{\delta }]=N-\frac{D}{2},
\end{equation*}
where, we recall, $D$ is the dimension of space--time.

The relation between $\Delta $ and $\overline{\Delta }$ requires, on the
other hand, a very long a quite technical analysis, which we leave to the
appendix. Fortunately, though, the answer is very simple, and completely
parallel to the above results. In fact, as shown in Lemma \ref{L-7}, one has
that 
\begin{equation*}
\left[ \overline{\Delta },\Delta \right] =N-\frac{D}{2}.
\end{equation*}
We see that the structure of the seemingly different pairs of operators $%
\overline{\Delta },\Delta $ and $\delta ,\overline{\delta }$ is actually
very similar and compatible, and we will use the above results heavily in
the next section to solve the invariance equations for the simple case $D=2$.

Let us conclude this section by discussing the general form of the solution
of equations (\ref{eqinv1}) and (\ref{eqinv2}). We introduce the concept of 
\textit{lowest level tensor}, by which we mean a CGI tensor $\rho $ which
satisfies 
\begin{equation}
\overline{\Delta }\rho =\delta \rho =0.  \label{eqLLT}
\end{equation}
A general solution of (\ref{eqinv1}) and (\ref{eqinv2}) will then consist of
a lowest level tensor $\tau ^{P}$ at level $P$, together with tensors $\tau
^{L}$ at higher levels $L>P$, which are required in order to obtain an
invariant action. At each level $L$, $\tau ^{L}$ is determined using (\ref
{eqinv1}) and (\ref{eqinv2}) in terms of the tensors of lower level, up to
an addition $\tau ^{L}\rightarrow \tau ^{L}+\rho $ of a lowest level tensor $%
\rho $. Let us suppose that we can, given a $\tau ^{P}$ satisfying (\ref
{eqLLT}), construct, in a \textit{canonical }way, a tower of tensors $\tau
^{L}$, $L>P$ so that (\ref{eqinv1}) and (\ref{eqinv2}) hold. We will then
call the full set $\left\{ \tau ^{L}\right\} _{L\geq P}$ the \textit{%
invariant block} generated by $\tau ^{P}$. The above discussion then shows
that a solution of (\ref{eqinv1}) and (\ref{eqinv2}) is, in general, a
linear combination of invariant blocks. We will see in the next section
that, in the simple case $D=2$, we will indeed be able to construct
canonically invariant blocks starting from generic lowest level
tensors.\bigskip \newline
N{\small OTE. Recall, from the example in section \ref{bict}, \ that }$\tau
^{2}${\small \ is given by equation (\ref{lala}) and satisfies }$\overline{%
\Delta }\tau ^{2}=0${\small . It is trivial to check that also }$\delta \tau
^{2}=0${\small \ holds, since }$\tau ^{2}${\small \ depends only on the
symmetric part of }$\gamma _{ab}${\small . Therefore }$\tau ^{2}${\small \
is a lowest level state, as is natural to expect.\bigskip }

To conclude, let us comment on the question of uniqueness of the solution of
the recursion equations (\ref{eqinv1}) and (\ref{eqinv2}). As we just
discussed, we do not expect the solution to be unique, since general
solutions are in one to one correspondence with lowest level tensors. An
appropriate way of thinking about (\ref{eqinv1}) and (\ref{eqinv2}) is
probably by analogy with general relativity. In that case, one is free to
write actions of different type, subject only to the general principle of
covariance under diffeomorphisms on the underlying space--time manifold. In
a similar way, equations (\ref{eqinv1}) and (\ref{eqinv2}) imply that the
action not only must be invariant under reparametrizations of the
world--volume of the brane, but must also be invariant under a more general
set of transformations, parametrized by changes in $\theta $.\ It would then
be of great practical importance to have an explicitly covariant notation,
for which invariance under (\ref{eqinv1}) and (\ref{eqinv2}) is manifest.

\subsection{\label{summ}Summary of the basic results}

We now summarize, in a compact but self--contained way, the results of this
paper. Invariant actions are described by tensors $\tau ^{L}$ for $L\geq 2$
such that

\begin{enumerate}
\item  The tensor $\tau ^{L}$ has $2L$ indices and is built from a basic
matrix $\gamma _{ab}$.

\item  The tensors $\tau ^{L}$ are cyclic gauge invariant -- \textit{i.e.}
are cyclic tensors which satisfy the algebraic relation 
\begin{equation*}
\tau _{123\cdots n}+\tau _{213\cdots n}+\tau _{231\cdots n}+\cdots +\tau
_{23\cdots 1n}=0.
\end{equation*}

\item  For any choice of antisymmetric matrix $\Delta _{ab}$ (with $\Delta
^{ab}$ indicating the inverse of $\Delta _{ab}$) the tensors $\tau ^{L}$
satisfy the basic relation 
\begin{eqnarray*}
\Delta \tau ^{L} &=&\delta \tau ^{L+1} \\
\overline{\Delta }\tau ^{L+1} &=&\overline{\delta }\tau ^{L}
\end{eqnarray*}
where the differential operators $\delta ,\overline{\delta }$ are given by 
\begin{eqnarray*}
\delta &=&\gamma _{ab}\frac{\partial }{\partial \gamma _{ab}} \\
\overline{\delta } &=&\frac{1}{2}\left( \gamma _{ab}\Delta ^{ab}\right)
+\left( \gamma \Delta \gamma \right) _{ab}\frac{\partial }{\partial \gamma
_{ab}}
\end{eqnarray*}
and the algebraic operators $\Delta ,\overline{\Delta }$ are given by 
\begin{equation*}
\left( \Delta \tau \right) _{1\cdots n}=\frac{i}{2}\left( \frac{n-2}{n}%
\right) \left( \Delta _{12}\tau _{345\cdots n}-\Delta _{13}\tau _{245\cdots
n}\right) +\mathrm{cyc}_{1\cdots n}
\end{equation*}
and 
\begin{equation*}
\left( \overline{\Delta }\tau \right) _{1\cdots n}=-\frac{i}{2}\left( \frac{%
n+2}{n^{2}}\right) \Delta ^{ab}\left[ n\tau _{1\cdots nab}+\left( n-1\right)
\tau _{1\cdots anb}+\cdots +0\cdot \tau _{a1\cdots nb}\right] +\mathrm{cyc}%
_{1\cdots n}
\end{equation*}
The operators $\Delta ,\overline{\Delta },\delta ,\overline{\delta }$
satisfy the commutation relations 
\begin{equation*}
\left[ \overline{\Delta },\Delta \right] \tau ^{L}=\left[ \delta ,\overline{%
\delta }\right] \tau ^{L}=\left( 2L-\frac{D}{2}\right) \tau ^{L}.
\end{equation*}
\end{enumerate}

\section{\label{S6}A Solution in $2$ Dimensions}

In this section we describe a general constructive solution of the
invariance equations (\ref{eqinv1}) and (\ref{eqinv2}) in dimension $D=2$.
This is clearly a toy model, since gauge fields have no propagating degrees
of freedom in $2$ dimensions. On the other hand the solution is still highly
non--trivial, and it exhibits many of the features which are expected to be
present in the general $D$--dimensional case.

\subsection{The general strategy}

It is convenient, in two dimensions, to use on $M$ complex coordinates 
\begin{equation*}
z=x^{1}+ix^{2}\;\ \ \ \ \ \ \ \ \ \ \ \ \ \overline{z}=x^{1}-ix^{2}
\end{equation*}
with a hermitian metric 
\begin{equation*}
g_{z\overline{z}}=g
\end{equation*}
and $g_{zz}=g_{\overline{z}\overline{z}}=0$. Similarly, the antisymmetric
tensors $B_{ab}$ and $K_{ab}$ have a single independent component 
\begin{equation*}
B_{z\overline{z}}=iB\,\ \ \ \ \ \ \ \ \ \ \ \ \ \ \ \ \ \ \ K_{z\overline{z}%
}=iK
\end{equation*}
The tensor $\gamma _{ab}$ is then given by a single complex number 
\begin{eqnarray*}
\gamma _{z\overline{z}} &=&g+i(B-K)=x+iy \\
\gamma _{\overline{z}z}. &=&\overline{\gamma _{z\overline{z}}}=x-iy
\end{eqnarray*}
Finally we will use $Z,\overline{Z}$ for the operators which correspond to
the coordinates $z,\overline{z}$ under the map $Q_{\omega }$.

We adopt, in this section, a notation which is not well suited for the
general $D$--dimensional case treated in the remainder of the paper, but
which is more economical in the present setting. Consider a general term of
level $L$ in the action 
\begin{equation*}
\tau _{1\cdots 2L}\mathrm{Tr}\left( X^{1}\cdots X^{2L}\right)
\end{equation*}
and let the various indices $1,2,\cdots ,2L$ run over their possible values $%
z,\overline{z}$. We obtain a sum consisting of traces of monomials in $Z,%
\overline{Z}$, multiplied by polynomials of degree $L$ in $x,y$. In
particular, the monomials under the trace satisfy the following two
properties

\begin{enumerate}
\item  They are constructed with $L$ coordinates $Z$ and $L$ coordinates $%
\overline{Z}$.

\item  Monomials which differ only by a cyclic permutation of the coordinate
operators $Z,\overline{Z}$ should be considered, as we recall from section 
\ref{bict}, as identical.
\end{enumerate}

We call objects which satisfy ($1$) and ($2$)\thinspace cyclic words of
level $L$ -- or simply words -- and we will denote with $W_{L}$ the space of
their linear combinations (for example, for $L=2$, the space $W_{2}$ is
spanned by the two words $ZZ\overline{Z}\overline{Z}$ and $Z\overline{Z}Z%
\overline{Z}$). Also we let $P_{L}$ be the space of polynomials in $x,y$ of
degree $L$.

Among the possible cyclic words in $W_{L}$, we must consider the subspace 
\begin{equation*}
G_{L}\subset W_{L}
\end{equation*}
of cyclic gauge invariant combinations. Following once more the discussion
of section \ref{bict}, we define the space $G_{L}$ as follows. First
introduce canonical creation and annihilation operators $a$ and $a^{\dagger
} $, which satisfy $[a,a^{\dagger }]=1$, and let $O$ be the space of
polynomials in $a,a^{\dagger }$. Consider then a map 
\begin{equation*}
r:W_{L}\rightarrow O
\end{equation*}
defined by taking a word $w$ in $W_{L}$ and by constructing the operator $%
r(w)$ by replacing $Z,\overline{Z}$ with $a,a^{\dagger }$, and then by
summing over the possible cyclic permutations. For example, we associate to
the word $w=\overline{Z}\overline{Z}ZZ$ the operator\footnote{%
Note that the sum over cyclic permutations is crucial in order to have a
well--defined map $r$, since $W_{L}$ consists of words defined only up to
cyclic permutation of the letters $Z,\overline{Z}$. In the above example,
the same word $w$ can be equally represented by any of the permutations $w=%
\overline{Z}\overline{Z}ZZ=\overline{Z}ZZ\overline{Z}=ZZ\overline{Z}%
\overline{Z}=Z\overline{Z}\overline{Z}Z$, but the operator $r(w)$ is
independent of the choice of representative for $w$.} 
\begin{equation*}
r(w)=a^{\dagger }a^{\dagger }aa+aa^{\dagger }a^{\dagger }a+aaa^{\dagger
}a^{\dagger }+a^{\dagger }aaa^{\dagger }.
\end{equation*}
We then have that 
\begin{equation*}
G_{L}=\ker r.
\end{equation*}
This equation restates the fact (proven in Lemma \ref{L-1}) that, given a
cyclic gauge invariant tensors $\tau _{1\cdots 2L}$, the function $\left(
x^{1}\star _{\omega }\cdots \star _{\omega }x^{2L}\right) \tau _{1\cdots 2L}$
vanishes whenever $\omega =K$ is constant.

The operators $\delta ,\overline{\delta },\Delta ,\overline{\Delta }$ act
naturally on the spaces $P_{L}$ and $G_{L}$. In particular the operators $%
\delta $ and $\overline{\delta }$ act on the spaces $P_{L}$ of polynomials 
\begin{eqnarray*}
\delta &:&P_{L}\rightarrow P_{L+1} \\
\overline{\delta } &:&P_{L}\rightarrow P_{L-1}
\end{eqnarray*}
Choosing, without loss in generality, $\Delta _{z\overline{z}}=\Delta ^{z%
\overline{z}}=i$, we have 
\begin{eqnarray*}
\delta &=&\frac{\partial }{\partial y} \\
\overline{\delta } &=&-y+2xy\frac{\partial }{\partial x}-\left(
x^{2}-y^{2}\right) \frac{\partial }{\partial y}
\end{eqnarray*}
One can check explicitly that $[\delta ,\overline{\delta }]=2x\partial
_{x}+2y\partial _{y}-1=2L-1$.

The operators $\Delta $, $\overline{\Delta }$ on the other hand act on the
spaces $G^{L}$%
\begin{eqnarray*}
\Delta &:&G_{L}\rightarrow G_{L+1} \\
\overline{\Delta } &:&G_{L}\rightarrow G_{L-1}
\end{eqnarray*}
We will be more explicit on the precise form of $\Delta $ and $\overline{%
\Delta }$ in the next subsection, but we know, from the general arguments of
section \ref{iuctbc}, that$\,[\overline{\Delta },\Delta ]=2L-1$.

With this notation in place we can now easily construct invariant actions.
In particular, we will first show how to canonically construct, starting
from a lowest level term, a complete set of terms which combine into an
invariant block. A general invariant action is then given, following the
discussion at the end of section \ref{iuctbc}, by linear combinations of
invariant blocks.

Let us then first describe the form of a lowest level term. In general,
given the above discussion, a generic term in the action at level $L$ will
be of the form 
\begin{equation*}
S_{L}=\sum_{i}p_{L}^{i}g_{L}^{i}.
\end{equation*}
where the $g_{L}^{i}$ are a basis for $G_{L}$ and the $p_{L}^{i}$ are
polynomials in $P_{L}$. A term $S_{P}=\sum_{i}p_{P}^{i}g_{P}^{i}$ (we will
reserve $L$ for a general level index, and $P$ for lowest level states) will
be of lowest level if 
\begin{equation}
\delta S_{P}=0\,\ \ \ \ \ \ \ \ \ \ \ \ \ \ \ \ \ \ \ \overline{\Delta }%
S_{P}=0  \label{eq60-100}
\end{equation}
The first equation implies that the polynomials $p_{P}^{i}$ depend uniquely
on $x$, and therefore that $p_{P}^{i}=c_{i}x^{P}$. Then 
\begin{equation*}
S_{P}=x^{P}g_{P}
\end{equation*}
where $g_{P}=\sum_{i}c_{i}g_{P}^{i}$ satisfies, using the second equation in
(\ref{eq60-100}), 
\begin{equation*}
\overline{\Delta }g_{P}=0.
\end{equation*}
Let us then start with a lowest level term $S_{P}$ and construct a full
invariant block $S$ of the form 
\begin{eqnarray*}
S &=&\sum_{L\geq P}S_{L} \\
S_{L} &=&p_{L}g_{L}.
\end{eqnarray*}
The above is invariant if 
\begin{eqnarray}
\Delta S_{L} &=&\delta S_{L+1}  \label{pippo} \\
\overline{\delta }S_{L} &=&\overline{\Delta }S_{L+1}  \notag
\end{eqnarray}
To solve the above constraints we construct the higher level $g_{L}$'s using 
$\Delta $%
\begin{equation*}
g_{L+1}=\Delta g_{L}
\end{equation*}
We must then find polynomials $p_{L}$ which satisfy 
\begin{equation}
\delta p_{L+1}=p_{L}  \label{ubababa}
\end{equation}
and such that 
\begin{eqnarray}
\overline{\Delta }g_{L+1} &=&c_{L}g_{L}  \notag \\
\overline{\delta }p_{L} &=&c_{L}p_{L+1}  \label{ubauba}
\end{eqnarray}
for some constant $c_{L}$. First we compute $\overline{\Delta }g_{L+1}$.
Using the fact that $[\overline{\Delta },\Delta ]=2L-1$, and that $\overline{%
\Delta }g_{P}=0$, we obtain 
\begin{eqnarray*}
\overline{\Delta }g_{L+1} &=&\overline{\Delta }\Delta ^{L+1-P}g_{P}=\left(
\left( 2L-1\right) +\cdots +(2P-1)\right) g_{L} \\
&=&c_{L}g_{L}
\end{eqnarray*}
with 
\begin{equation*}
c_{L}=(L+P-1)(L-P+1).
\end{equation*}
Using equation (\ref{ubauba}) 
\begin{equation}
p_{L+1}=\frac{1}{(L+P-1)(L-P+1)}\overline{\delta }p_{L}  \label{tubatuba}
\end{equation}
to define higher level polynomials, we can check, using $[\delta ,\overline{%
\delta }]=2L-1$ and $\delta p_{P}=0$, that the remaining equation (\ref
{ubababa}) is satisfied, and that we have indeed a solution to the
invariance equations.

\subsection{\label{ie2d}An important example}

In this subsection we use the general construction described above and apply
it to a specific important example. In particular we show again that the
basic $F^{2}$ term is lowest level, and we construct part of the invariant
block constructed from it. We recover in particular the Born--Infeld action,
and we compute the first non--trivial derivative corrections at level $4$
which must be present in order to make the full action invariant.

We start by analyzing the explicit form of the operators $\Delta ,\overline{%
\Delta }$, recalling that $\Delta _{z\overline{z}}=\Delta ^{z\overline{z}}=i$%
. We define for convenience the \textit{field strength} $F$ as 
\begin{equation*}
F=\left[ Z,\overline{Z}\right] .
\end{equation*}
Then equation (\ref{eq40-100}) reads, in the present case, 
\begin{eqnarray*}
\Delta Z &=&\frac{1}{4}(ZF+FZ) \\
\Delta \overline{Z} &=&\frac{1}{4}(\overline{Z}F+F\overline{Z}).
\end{eqnarray*}
The above then defines the action of $\Delta $ on $G_{L}$, since $\Delta $
acts as a derivation on each coordinate forming the words in $G_{L}$.
Similarly $\overline{\Delta }$ is defined by 
\begin{equation*}
\overline{\Delta }F=1.
\end{equation*}
More precisely, given a word $w\in W_{L}$, we cyclically rearrange the
coordinates in each word so as to obtain a gauge invariant (\textit{not}
cyclic gauge invariant) form, containing only commutators. We then apply $%
\overline{\Delta }$ on each fundamental commutator $F$ as a derivation.

Let us the consider the term 
\begin{equation*}
S_{2}=p_{2}g_{2}
\end{equation*}
with 
\begin{equation*}
p_{2}=x^{2}\,\ \ \ \ \ \ \ \ \ \ \ \ \ \ \ \ \ \ \ \ \ \ \ g_{2}=\frac{1}{2}%
F^{2}.
\end{equation*}
Clearly $\delta p_{2}=0$. Moreover $\overline{\Delta }g_{2}=F=0$, since we
recall that, as a word in $W_{1}$, the commutator $F$ is zero. Therefore $%
S_{2}$ is a lowest level state, and we can construct the corresponding
invariant block.

First we do some computations explicitly. The polynomials $p_{3}$ and $p_{4}$
are given by 
\begin{eqnarray*}
p_{3} &=&\frac{1}{3}\overline{\delta }p_{2}=yx^{2} \\
p_{4} &=&\frac{1}{8}\overline{\delta }p_{3}=-\frac{1}{8}x^{4}+\frac{1}{2}%
y^{2}x^{2}.
\end{eqnarray*}
We then compute $g_{3}$ and $g_{4}$. First, applying the operation $\Delta $
to $F$ we obtain 
\begin{equation*}
\Delta F=\frac{1}{2}\left( F^{2}+ZF\overline{Z}-\overline{Z}FZ\right) .
\end{equation*}
This means that (recall that the RHS below is a word in $W_{3}$, and that
cyclic rearrangements are allowed) 
\begin{eqnarray*}
g_{3} &=&\Delta g_{2}=\frac{1}{2}F\left( \Delta F\right) +\frac{1}{2}\left(
\Delta F\right) F=F\left( \Delta F\right) \\
&=&\frac{1}{2}F^{3}+\frac{1}{2}[FZ,F\overline{Z}]=\frac{1}{2}F^{3}
\end{eqnarray*}
The computation of $g_{4}$ is just slightly more complex, and we leave it
for the appendix (Lemma \ref{L-10}). The result is 
\begin{equation*}
g_{4}=\Delta g_{3}=F^{4}+\frac{1}{4}F[DF,\overline{D}F]
\end{equation*}
where 
\begin{equation*}
D\cdots =[Z,\cdots ]\,\ \ \ \ \ \ \ \ \ \ \ \ \ \ \ \ \ \overline{D}\cdots =[%
\overline{Z},\cdots ].
\end{equation*}
We can now combine the polynomials and the gauge invariant words. To make
contact with standard notation we write the action for $y=0$ ($\gamma _{ab}$
symmetric) and revert to more standard notation 
\begin{eqnarray}
F &\rightarrow &iF_{z\overline{z}}=iF  \label{eqsuca} \\
x &\rightarrow &g^{z\overline{z}}=\frac{1}{g}  \notag \\
D &\rightarrow &iD  \notag
\end{eqnarray}
which give the following $U(N)\,$\ lagrangian 
\begin{equation}
-\frac{1}{2g^{2}}\mathrm{Tr}\left( F^{2}\right) -\frac{1}{8g^{4}}\mathrm{Tr}%
\left( F^{4}+\frac{i}{2}F[DF,\overline{D}F]\right) +\cdots .
\label{abracadabra}
\end{equation}
We see that the above action, written up to level $4$, contains the first
part of the Born--Infeld action (in $2$--dimensions there is no ambiguity
about ordering of the $F^{2n}$ terms), but already at level $4$ we have
derivative corrections, which are required for the total invariance of the
action\footnote{%
It has been shown in \cite{TF4} that, at level $4$, the effective action can
be written only as the $F^{4}$ term, with no derivative corrections. This
result is not in contradiction with equation (\ref{abracadabra}), since one
can always allow for field redefinitions. In particular, consider, in
general $D$ dimensions, the field redefinition $A_{a}\rightarrow
A_{a}+cF_{ab}D_{c}F_{bc}$. This induces a change in the action at level $4$,
coming from the $F^{2}$ term, of the form $F_{da}D_{d}\left(
F_{ab}D_{c}F_{bc}\right) =\frac{1}{2}F_{ab}\left[ D_{c}F_{bc},D_{d}F_{ad}%
\right] $, which in two dimensions is proportional to $F[DF,\overline{D}F]$.
Therefore, for an appropriate choice of $c$, one can remove the derivative
term in (\ref{abracadabra}), thus resolving the apparent contradiction with 
\cite{TF4}.}.

Let us now schematically consider the higher terms $g_{L}$. We wish to
sketch how one can recover, within the $F^{2}$ invariant block, the complete
Born--Infeld action (a much more detailed discussion on this point and
related issues will appear in \cite{TwoT}). To this end, we first note that,
in general, 
\begin{equation*}
g_{L}=c_{L}F^{L}+\mathrm{derivative\;terms.}
\end{equation*}
We have that $c_{2}=c_{3}=1/2$, $c_{4}=1$,$\cdots $. We can use the basic
commutator $\left[ \overline{\Delta },\Delta \right] $ and the fact that $%
\overline{\Delta }F^{L}=LF^{L-1}$ to compute all the $c_{L}$'s. In fact,
applying the basic equation $\left[ \overline{\Delta },\Delta \right] =2L-1$
to $g_{L}$ (recalling that $\Delta g_{L}=g_{L+1}$ and that $\overline{\Delta 
}$ does not decrease the number of derivatives) one obtains the recursion
relation 
\begin{equation*}
\left( L+1\right) \left( \frac{c_{L+1}}{c_{L}}\right) -L\left( \frac{c_{L}}{%
c_{L-1}}\right) =2L-1
\end{equation*}
which is solved by $c_{L}=(L-2)c_{L-1}$, or by 
\begin{equation*}
c_{L}=\frac{1}{2}(L-2)!
\end{equation*}
The $F^{2}$ invariant block then contains the sum $%
\sum_{L}c_{L}p_{L}(x,y)F^{L}$. Let us consider more closely the polynomials $%
p_{L}$. First of all, from the general relation (\ref{tubatuba}) one can
easily show that the polynomials $p_{L}$ vanish for $L$ odd if $y=0$. On the
other hand, for even levels, one has that 
\begin{equation*}
p_{2L}\left( x,0\right) =d_{2L}x^{2L}.
\end{equation*}
Therefore, the relevant part of the action is given by (again substituting $%
x\rightarrow 1/g$ and $F\rightarrow iF$) 
\begin{equation}
\sum_{L}c_{2L}d_{2L}\frac{\left( -\right) ^{L}}{g^{2L}}\mathrm{Tr}\left(
F^{2L}\right) .  \label{arfarf}
\end{equation}
To compute the coefficients $d_{2L}$, let us first note that, since $\delta
p_{2L-1}=p_{2L-2}$, one must have that $p_{2L-1}=d_{2L-2}x^{2L-2}y+o\left(
y^{3}\right) $. This implies, using (\ref{tubatuba}), that$\;d_{2L}=-\frac{1%
}{4L\left( L-1\right) }d_{2L-2}$, which is solved by (recall that $d_{2}=1$) 
\begin{equation*}
d_{2L}=\left( -\frac{1}{4}\right) ^{L-1}\frac{1}{L!\left( L-1\right) !}.
\end{equation*}
Therefore, equation (\ref{arfarf}) gives the complete Born--Infeld action 
\begin{equation*}
-\sum_{L}\left( \frac{1}{2}\right) ^{2L-1}\frac{(2L-2)!}{L!\left( L-1\right)
!}\frac{1}{g^{2L}}\mathrm{Tr}\left( F^{2L}\right) =\mathrm{Tr}\sqrt{1-\frac{1%
}{g^{2}}F^{2}}.
\end{equation*}

\section{\label{S7}Discussion}

In this paper we have analyzed in detail the general structure of the
non--abelian Born--Infeld action, together with the higher $\alpha ^{\prime
} $ derivative corrections. We have shown how the requirement of invariance
of the action under a change of non--commutativity scale $\theta $ imposes
severe restrictions on the possible terms which can appear. More
specifically, we can construct invariant actions starting from invariant
blocks, which are themselves obtained from a lowest level term (in a loose
sense, a pure derivative term). Terms at higher level are then constructed
so as to achieve invariance under a change in $\theta $. A general action is
then a linear combination of invariant blocks, with coefficients which must
be determined from a different computation. No argument in this paper
assumes supersymmetry, and the results are therefore valid in bosonic open
string theory, as well as in superstring theory. In particular,
supersymmetry will impose restrictions on the allowed linear combinations of
invariant blocks, possibly determining in part, or even completely, the
effective action.

Let us now comment on interesting directions of possible future
investigation.

\begin{itemize}
\item  It is first of all important to explicitly solve the invariance
equations $\Delta \tau ^{L}=\delta \tau ^{L+1}$ and $\overline{\Delta }\tau
^{L+1}=\overline{\delta }\tau ^{L}$ in the general $D$--dimensional case.
Similarly to the $2$--dimensional case discussed in the text, we should
study the algebra of operators $\Delta ,\delta ,\overline{\Delta },\overline{%
\delta }$ given by the relations $\left[ \overline{\Delta },\Delta \right]
=[\delta ,\overline{\delta }]=2L-D/2$. The algebra now depends on more
parameters, since the underlying matrix $\Delta _{ab}$ now has $D\left(
D-1\right) /2$ components. It is important, in particular, to have a
canonical construction of higher level terms, starting from the lowest
level. This would in turn give a canonical definition of invariant block.

\item  It is important to understand how invariant blocks appear in the
underlying boundary conformal field theory. In particular, the relation
between the analysis of this paper, which is at the level of the effective
action, and conformal field theory is of importance both conceptually and
from a practical point of view.

\item  The results of this paper do not depend on supersymmetry.
Understanding the additional constraints imposed by SUSY is an important
task for the future. A first step in this direction is the following. Given
an invariant action, we may use T--duality to describe the weak--coupling
physics of D--branes. In particular, we expect, in a supersymmetric theory,
to have minima of the effective action corresponding to holomorphic curves,
surfaces, $\cdots $. Very possibly, a careful restatement of this fact in
terms of invariant blocks will impose constraints which must be satisfied in
a supersymmetric theory.

\item  Given the invariant description of the action (\ref{eq50-200}) as an
operator trace, it is very tempting to resum the full series in one specific
invariant block. In fact, although the action is usually written by
artificially choosing a parameter $\theta $ and then by writing the
expression in terms of coordinate operators $X^{a}$, it is nonetheless true
that the operator $O(\theta )=\sum_{L}X^{1}\cdots X^{2L}\tau _{1\cdots 2L}$
has a trace $\mathrm{Tr}\left( O\left( \theta \right) \right) $ which is $%
\theta $--independent. It is then tempting to conjecture\footnote{%
The spectral nature of actions has been very much stressed by A.Connes.}
that the various operators $O(\theta )$ not only have the same trace, but
are related by a unitary transformation, and are then isospectral.
\end{itemize}

\section{Acknowledgments}

I would like to thank M. Douglas, A. Connes, A. Cattaneo and B. Shoikhet for
useful discussions. Also I want to thank the Institute for Advanced Study
and MIT, and in particular N. Seiberg and W. Taylor, for hospitality during
the completion of this work. This research is supported by a European
Post--doctoral Institute Fellowship.

\section{Appendix}

\begin{definition}
A tensor $\eta _{1\cdots n}$ is called gauge invariant (GI) if 
\begin{equation*}
\eta _{123\cdots n}+\eta _{213\cdots n}+\eta _{231\cdots n}+\cdots +\eta
_{23\cdots 1n}+\eta _{23\cdots n1}=0.
\end{equation*}
A tensor $\tau _{1\cdots n}$ is called cyclic gauge invariant (CGI)\ if 
\begin{equation}
\tau _{123\cdots n}+\tau _{213\cdots n}+\tau _{231\cdots n}+\cdots +\tau
_{23\cdots 1n}=0.  \label{A1}
\end{equation}
\end{definition}

\begin{lemma}
(Section \ref{bict})\label{L-cgi} Let $\eta _{1\cdots n}$ be gauge
invariant. Then 
\begin{equation*}
\tau _{1\cdots n}=\frac{1}{n}\left( \eta _{1\cdots n}+\mathrm{cyc}_{1\cdots
n}\right)
\end{equation*}
is cyclic gauge invariant.
\end{lemma}

P{\small ROOF}. Let us write the left hand side of equation (\ref{A1}) in
terms of $\eta $. Neglecting the multiplicative factor of $1/n$, we have the
expression 
\begin{eqnarray*}
&&\eta _{123\cdots n}+\eta _{213\cdots n}+\eta _{231\cdots n}+\cdots +\eta
_{23\cdots 1n}+ \\
&&\eta _{23\cdots n1}+\eta _{13\cdots n2}+\eta _{31\cdots n2}+\cdots +\eta
_{3\cdots 1n2}+ \\
&&\eta _{3\cdots n12}+\eta _{3\cdots n21}+\cdots \\
&&\cdots \\
&&\cdots +\eta _{1n23\cdots n-1}+ \\
&&\eta _{n123\cdots n-1}+\eta _{n213\cdots n-1}+\eta _{n231\cdots
n-1}+\cdots +\eta _{n23\cdots n-1,1}
\end{eqnarray*}
The above expression contains $\left( n-1\right) n$ terms ($n$ lines with $%
n-1$ terms each). Consider the sequence of terms in the order written, and
assemble them now into groups of $n$ terms. It is then easy to see that each
individual group vanishes since $\eta $ is gauge invariant. $\square $

\begin{lemma}
(Section \ref{bict}) \label{L-1}Let $\tau _{1\cdots n}$ be cyclic gauge
invariant and let 
\begin{equation*}
\tau (x)=x^{1}\star _{\omega }\cdots \star _{\omega }x^{n}\;\tau _{1\cdots
n}.
\end{equation*}
Then, for $x\rightarrow \infty $, $\tau (x)$ grows linearly with $x$.
Moreover, if $n$ is even, then 
\begin{equation*}
\tau (x)\rightarrow 0
\end{equation*}
for $x\rightarrow \infty .$
\end{lemma}

P{\small ROOF}. We use the fact that, at infinity, $\omega \rightarrow K$
approaches a constant, and that we can therefore compute the variation 
\begin{eqnarray*}
\tau (x+\varepsilon )-\tau (x) &=&\varepsilon ^{1}x^{2}\star _{K}\cdots
\star _{K}x^{n}(\tau _{12\cdots n}+\cdots +\tau _{2\cdots n1}) \\
&=&h(x),
\end{eqnarray*}
where 
\begin{equation*}
h(x)=x^{2}\star _{K}\cdots \star _{K}x^{n}\,\ \varepsilon ^{1}\tau
_{12\cdots n}.
\end{equation*}
It is immediate to see that, for any $\varepsilon $, the tensor $\varepsilon
^{1}\tau _{12\cdots n}$ is gauge invariant in the indices $2,\cdots ,n$ and
that, using the results of section \ref{bict}, the function $h(x)$ is
constant at infinity. This implies that $\tau (x)$ is at most a linear
function of the coordinates $x$, when $x\rightarrow \infty $. We now note
that $x^{1}\star _{K}\cdots \star _{K}x^{n}$ is a polynomial of degree $n$,
with monomials of degrees $n,n-2,n-4,\cdots $. In particular, if $n$ is
even, we do not have a linear term and the function $\tau $ approaches a
constant at infinity. We now wish to show that the constant is $0$. Consider
the polynomial $x^{1}\star _{K}\cdots \star _{K}x^{n}$. It will be of the
form 
\begin{equation*}
x^{1}\star _{K}\cdots \star _{K}x^{n}=C^{1\cdots n}+o(x).
\end{equation*}
We have just seen that we do not need to consider the $o(x)$ part, and we
therefore just need to prove that $C^{1\cdots n}\tau _{1\cdots n}=0$. It is
not difficult to show, using the Moyal product $\star _{K}$, that 
\begin{equation}
C^{1\cdots n}\varpropto \sum_{\sigma }(-)^{J(\sigma )}\theta ^{\sigma
_{1}\sigma _{2}}\cdots \theta ^{\sigma _{n-1}\sigma _{n}}  \label{AAA234}
\end{equation}
where $J(\sigma )$ counts the number of pairs $\sigma _{2i-1},\sigma _{2i}$
for which $\sigma _{2i-1}>\sigma _{2i}$. Then, in order to finish the proof,
one has to show that the quantity 
\begin{equation}
\sum_{\sigma }(-)^{J(\sigma ^{-1})}\tau _{\sigma _{1}\cdots \sigma _{n}}
\label{A2}
\end{equation}
vanishes for cyclic $\tau $'s. Let us then fix a given permutation $\sigma $%
, and let me denote with $\pi $ the basic cyclic permutation $\left(
1,\cdots ,n\right) \rightarrow \left( 2,\cdots ,n,1\right) $. Consider then
the permutations $\rho _{k}=\pi ^{k}\circ \sigma $. Let us first show that $%
(-)^{J(\rho _{k}^{-1})}$ is alternating with $k$. In fact, for $k\rightarrow
k+1$, almost all the pairs $\sigma _{2i-1},\sigma _{2i}$ go into the pairs $%
\sigma _{2i-1}+1,\sigma _{2i}+1$. This is, on the other hand, not true for
the single pair with either $\sigma _{2i-1}$ or $\sigma _{2i}$ equal to $n$,
since $n\rightarrow 1$. Only this one pair changes the ordering of its
components, and therefore the sign $(-)^{J(\rho _{k}^{-1})}$ changes if $%
k\rightarrow k+1$. Consider then the set of permutations $\rho _{k}$, cyclic
permutations of $\sigma $, for $0\leq k<n$. This gives, in the sum (\ref{A2}%
), 
\begin{equation*}
\pm \left( \tau _{\sigma _{1}\sigma _{2}\sigma _{3}\cdots \sigma _{n}}-\tau
_{\sigma _{2}\sigma _{3}\cdots \sigma _{n}\sigma _{1}}+\tau _{\sigma
_{3}\cdots \sigma _{n}\sigma _{1}\sigma _{2}}-\cdots -\tau _{\sigma
_{n}\sigma _{1}\sigma _{2}\cdots \sigma _{n-1}}\right)
\end{equation*}
The terms come with alternating signs, and since $n$ is even the number of $%
+ $ signs is equal to that of $-$ signs. Moreover, all the terms are
actually the same, since $\tau $ is cyclically symmetric. The above sum then
vanishes. Partitioning the set of all permutations $\sigma $ in sets of
cyclically related permutations, we can then show that the full sum (\ref{A2}%
) vanishes. $\square $

\begin{remark}
\label{RRR}(Lemmata \ref{L-2} and \ref{L-6}) We make a general comment on
integration of commutators, which will be useful in the rest of the
appendix. Consider two functions $f$ and $g$, and look at the integral 
\begin{equation*}
\int d^{D}x\,V(\omega )\;\left( f\star _{\omega }g-g\star _{\omega }f\right)
.\,
\end{equation*}
It $f$ and $g$ vanish at infinity, then the above integral vanishes, as was
discussed in the main text. If, on the other hand, $f$ and $g$ do not go to
zero for $x\rightarrow \infty $, we can proceed as follows. Assume that $%
\omega =K$ outside of a compact domain $D\subset M$ and consider the
integral of $[f,g]_{\omega }$ over $D$ 
\begin{equation}
I=\int_{D}d^{D}x\text{ }V(\omega )\text{ }[f,g]_{\omega }  \label{AAA10}
\end{equation}
If $f,g=0$ on $\partial D$ and outside of $D$, the above expression
vanishes, and therefore, in general, the integral (\ref{AAA10}) must reduce
to a boundary integral over $\partial D$. We can then continuously deform $%
\omega \rightarrow K$ in the interior of $D$ without changing the integral.
This means, in particular, that, for any functions $f$ and\thinspace $g$, 
\begin{equation*}
I=\int_{D}d^{D}x\text{ }\det {}^{%
{\frac12}%
}K\text{ }[f,g]_{K}.
\end{equation*}
From the above arguments it is also clear that the above equality holds even
if $f$ and $g$ depend themselves in a local way on $\omega $. For example,
if $f=f_{1}\star _{\omega }f_{2}$, then we have 
\begin{equation*}
I=\int_{D}d^{D}x\text{ }\det {}^{%
{\frac12}%
}K\text{ }[f_{1}\star _{K}f_{2},g]_{K}.
\end{equation*}
In practice, when integrating commutators, we can replace $\omega $ with $K$
in all the expressions without changing the integral.
\end{remark}

\begin{lemma}
\label{L-2}(Section \ref{bict}) Let $\eta _{1\cdots n}$ be gauge invariant
and let $\tau _{1\cdots n}=\frac{1}{n}\left( \eta _{1\cdots n}+\mathrm{cyc}%
_{1\cdots n}\right) $. Define 
\begin{eqnarray*}
\eta (x) &=&x^{1}\star _{\omega }\cdots \star _{\omega }x^{n}\;\eta
_{1\cdots n} \\
\tau (x) &=&x^{1}\star _{\omega }\cdots \star _{\omega }x^{n}\;\tau
_{1\cdots n}
\end{eqnarray*}
with $\eta (x)\rightarrow \eta _{\infty }$ for $x\rightarrow \infty $. If $n$
is even, then 
\begin{equation*}
\int d^{D}xV(\omega )\,\tau =\int d^{D}xV(\omega )\left[ \eta -\eta _{\infty
}\right]
\end{equation*}
\end{lemma}

P{\small ROOF}. First it is clear that 
\begin{eqnarray*}
\tau (x) &=&\frac{1}{n}\,\eta _{1\cdots n}(x^{1}\star _{\omega }\cdots \star
_{\omega }x^{n}+ \\
&&+x^{2}\star _{\omega }\cdots \star _{\omega }x^{n}\star _{\omega
}x^{1}+\cdots \\
&&+x^{n}\star _{\omega }x^{1}\star _{\omega }\cdots \star _{\omega }x^{n-1}).
\end{eqnarray*}
Therefore, the difference $\eta -\tau $ is given by 
\begin{eqnarray*}
\eta -\tau &=&\frac{1}{n}\,\eta _{1\cdots n}([x^{1},x^{2}\star _{\omega
}\cdots \star _{\omega }x^{n}]_{\omega } \\
&&+[x^{1}\star _{\omega }x^{2},x^{3}\star _{\omega }\cdots \star _{\omega
}x^{n}]_{\omega }+\cdots \\
&&+[x^{1}\star _{\omega }\cdots \star _{\omega }x^{n-1},x^{n}]_{\omega }).
\end{eqnarray*}
Recalling remark \ref{RRR}, we will be done once we show that the RHS above
is equal to 
\begin{equation*}
\eta _{\infty }=\eta _{1\cdots n}x^{1}\star _{K}\cdots \star _{K}x^{n}
\end{equation*}
when we replace $\omega \rightarrow K$. We must therefore prove that 
\begin{equation}
\eta _{\infty }=\frac{1}{n}\,\eta _{1\cdots n}([x^{1},x^{2}\star _{K}\cdots
\star _{K}x^{n}]_{K}+\cdots +[x^{1}\star _{K}\cdots \star
_{K}x^{n-1},x^{n}]_{K}).  \label{AAA456}
\end{equation}
Both the LHS and the RHS above are constant, since $\eta $ is GI. Let us
introduce a compact notation 
\begin{eqnarray*}
\eta _{1\cdots n}\,x^{1}\star _{K}\cdots \star _{K}x^{n} &\rightarrow
&[1\cdots n] \\
\eta _{1\cdots n}\,x^{2}\star _{K}\cdots \star _{K}x^{n}\star _{K}x^{1}
&\rightarrow &[2\cdots n1] \\
&&\cdots
\end{eqnarray*}
Using formula (\ref{AAA234}) and the arguments which follow it, we can show
that 
\begin{equation*}
\lbrack k\cdots n1\cdots k-1]=\left( -1\right) ^{k-1}[1\cdots n]=\left(
-1\right) ^{k-1}\eta _{\infty }.
\end{equation*}
Then the RHS of (\ref{AAA456}) is given by 
\begin{eqnarray*}
&&\frac{n-1}{n}[12\cdots n]-\frac{1}{n}[23\cdots n1]-\cdots -\frac{1}{n}[%
n1\cdots n-1] \\
&=&[12\cdots n]\left( \frac{n-1}{n}+\frac{1}{n}-\frac{1}{n}+\cdots +\frac{1}{%
n}\right) = \\
&=&[12\cdots n]
\end{eqnarray*}
as was to be shown. $\square $

\begin{lemma}
\label{L-5}(Section \ref{qmgc}) Let $f\,$\ be a generic function, and let $%
F=Q_{\omega }(f)$, where $\omega $ is an arbitrary symplectic structure. Let
also $\Delta _{ab}$ be a constant antisymmetric matrix. Then, to first order
in $\Delta $, \ 
\begin{equation*}
Q_{\omega +\Delta }=F+\frac{i}{4}\Delta _{ab}\left(
X^{a}X^{b}F+FX^{a}X^{b}-2X^{a}FX^{b}\right) .
\end{equation*}
\end{lemma}

P{\small ROOF}. We start by noting that, if $\omega =K$ is constant, a
simple computation using the Moyal product $\star _{K}$ shows that 
\begin{eqnarray}
f\star _{K+\Delta }g-f\star _{K}g &=&-\frac{i}{2}\left( \theta \Delta \theta
\right) ^{ab}\partial _{a}f\star _{K}\partial _{b}g=  \label{fig1} \\
&=&-\frac{i}{2}\Delta _{ab}\left[ x^{a},f\right] _{K}\star _{K}\left[ x^{b},g%
\right] _{K}.  \notag
\end{eqnarray}
We must then consider the product $f\star _{\omega +\Delta }g$ for general $%
\omega $. As always, we look for a map $T$ such that $T\left( f\star
_{\omega +\Delta }g\right) =Tf\star _{\omega }Tg$. If we work to first order
in $\Delta $, and accordingly let $T=1+R$ (with $R$ or order $\Delta $), one
has that 
\begin{equation}
f\star _{\omega +\Delta }g-f\star _{\omega }g=Rf\star _{\omega }g+f\star
_{\omega }Rg-R\left( f\star _{\omega }g\right) .  \label{fig2}
\end{equation}
The map $T=1+R$ also relates $Q_{\omega +\Delta }$ and $Q_{\omega }$ as
follows 
\begin{equation*}
Q_{\omega +\Delta }(f)=Q_{\omega }(f+Rf).
\end{equation*}
We have seen from the example in section \ref{s2de} that the we should
consider general variations $Q_{\omega +\Delta }(f)-Q_{\omega }(f)$ of the
form $\frac{i}{4}\Delta _{ab}\left(
aX^{a}X^{b}F+bFX^{a}X^{b}+cX^{a}FX^{b}\right) $ where $a,b,c$ are constants
which we must determine. It is then clear that 
\begin{equation*}
Rf=\frac{i}{4}\Delta _{ab}(ax^{a}\star x^{b}\star f+bf\star x^{a}\star
x^{b}+cx^{a}\star f\star x^{b}).
\end{equation*}
Using the above fact in the RHS of equation (\ref{fig2}), and comparing, for 
$\omega =K$, with the RHS of equation (\ref{fig1}), we obtain that $%
a=b=1,b=-2$, as was required. $\square $

\begin{lemma}
\label{L-6}(Section \ref{iuctbc}) Let $\tau _{1\cdots n-2}$ be a cyclic
gauge invariant tensor, and let $\Delta _{ab}$ be a constant antisymmetric
matrix. Then the tensor 
\begin{equation*}
\left( \Delta \tau \right) _{1\cdots n}=\frac{i}{2}\left( \frac{n-2}{n}%
\right) \left( \Delta _{12}\tau _{345\cdots n}-\Delta _{13}\tau _{245\cdots
n}\right) +\mathrm{cyc}_{1\cdots n}
\end{equation*}
is itself cyclic gauge invariant. Moreover, if $\omega $ is a generic
symplectic structure and $X^{a}=Q_{\omega }(x^{a})$, then, under the
variation $\omega \rightarrow \omega +\Delta $, the operator $\mathrm{Tr}%
\left( X^{1}\cdots X^{n-2}\right) \tau _{1\cdots n-2}$ varies by $\mathrm{Tr}%
\left( X^{1}\cdots X^{n}\right) \left( \Delta \tau \right) _{1\cdots n}$
whenever $n$ is even.
\end{lemma}

P{\small ROOF}. First we show that the tensor $\Delta \tau $ is indeed
cyclic gauge invariant. Written in full, we want to show that the following
sum 
\begin{eqnarray}
&&\left( \Delta _{12}\tau _{345\cdots n}-\Delta _{13}\tau _{245\cdots
n}\right) +\mathrm{cyc}_{123\cdots n}  \label{A1234} \\
&&\left( \Delta _{21}\tau _{345\cdots n}-\Delta _{23}\tau _{145\cdots
n}\right) +\mathrm{cyc}_{213\cdots n}  \notag \\
&&\cdots  \notag \\
&&\left( \Delta _{23}\tau _{45\cdots 1n}-\Delta _{24}\tau _{35\cdots
1n}\right) +\mathrm{cyc}_{23\cdots 1n}  \notag
\end{eqnarray}
vanishes. To this end, we consider three significant cases, with the hope
that the reader can understand from them the general line of the argument.

Let us first consider, within the above sum, terms which are proportional to 
$\Delta _{12}$. They come only from the first and the last line and are 
\begin{equation*}
\tau _{345\cdots n}-\tau _{n34\cdots n-1}=0.
\end{equation*}
Similarly, terms proportional to $\Delta _{21}$ come from the second and
third line and exactly cancel each other. Consider now terms proportional,
say, to $\Delta _{23}$. These terms are present in every line of the above
sum, and they are 
\begin{equation*}
\left( \tau _{145\cdots n}-\tau _{145\cdots n}\right) +(\tau _{145\cdots
n}+\tau _{415\cdots n}+\cdots +\tau _{45\cdots 1n})
\end{equation*}
The above vanishes since $\tau $ is cyclic gauge invariant. Finally we
consider terms which are proportional to $\Delta _{2n}$. In this final case,
we can check that no terms in the sum (\ref{A1234}) contain $\Delta _{2n}$.
The reader can convince him or herself that all other combination of indices
fall in one of these three cases.

\begin{remark}
Let us note that, in the above proof, we have not used the fact that $\Delta
_{ab}$ is antisymmetric. In fact, we have shown more generally that the
tensor 
\begin{equation*}
\left( A_{12}\tau _{345\cdots n}-A_{13}\tau _{245\cdots n}\right) +\mathrm{%
cyc}_{1\cdots n}
\end{equation*}
is cyclic gauge invariant for any choice of $A$, whenever $\tau $ itself is
CGI.
\end{remark}

We now move to the second part of the lemma. Introduce the following two
functions on $M$%
\begin{eqnarray*}
A(x) &=&x^{1}\star _{\omega }\cdots \star _{\omega }x^{n-2}\;\tau _{1\cdots
n-2} \\
B(x) &=&x^{1}\star _{\omega }\cdots \star _{\omega }x^{n}\;\left( \Delta
\tau \right) _{1\cdots n}.
\end{eqnarray*}
Since both $\tau $ and $\Delta \tau $ are CGI, the two functions $A$ and $B$
tend to $0$ as $x\rightarrow \infty $ (we are assuming $n$ even). Recall
that, under a variation $\omega \rightarrow \omega +\Delta $, the star
product of two functions $f\star _{\omega }g$ changes by $Rf\star g+f\star
Rg-R(f\star g)$ (we do not show the explicit $\omega $ dependence in $\star $%
), where $R$ is given in equation (\ref{eqa100}). Therefore, given three
functions, the variation of $f\star g\star h$ is $Rf\star g\star h+f\star
Rg\star h+f\star g\star Rh-R(f\star g\star h)$, and similarly for products
of more functions. In particular, the variation of the function $A$ is given
by 
\begin{equation}
A\rightarrow A+C-RA  \label{AA100}
\end{equation}
where 
\begin{equation*}
C(x)=Rx^{1}\star _{\omega }\cdots \star _{\omega }x^{n-2}\;\tau _{1\cdots
n-2}+\cdots +x^{1}\star _{\omega }\cdots \star _{\omega }Rx^{n-2}\;\tau
_{1\cdots n-2}.
\end{equation*}
Since we are interested in traces of operators, we must consider also the
variation coming from the change of integration measure. We use equations (%
\ref{eqa200}) and (\ref{AA100}), together with the fact that $A$ vanishes at
infinity, to show that the variation of 
\begin{equation*}
\mathrm{Tr}\left( X^{1}\cdots X^{n-2}\right) \tau _{1\cdots n-2}=\int
d^{D}x\,V(\omega )\,A
\end{equation*}
is simply given by 
\begin{equation*}
\int d^{D}x\,V(\omega )\,C.
\end{equation*}
We must then prove that 
\begin{equation*}
\int d^{D}x\,V(\omega )\,\left( B-C\right) =0.
\end{equation*}
It is simple to show that the function $B$ is obtained by cyclically
rearranging the coordinate functions which build $C$. Following remark \ref
{RRR}, the integral above reduces to a boundary term, and to show that it
vanishes we just need to prove that $B=C$ whenever $\omega =K$ is constant.
On one side, we know that $B=0$ for $\omega =K$, since $\Delta \tau $ is
cyclic gauge invariant. We then need to prove that, for constant symplectic
structures, $C=0$. This is shown in two steps. First look at the operation $%
R $ on coordinate functions in the case of flat symplectic structure 
\begin{eqnarray*}
Rx^{a} &=&\frac{i}{4}\Delta _{bc}\left( x^{b}\star _{K}x^{c}\star
_{K}x^{a}+x^{a}\star _{K}x^{b}\star _{K}x^{c}-2x^{b}\star _{K}x^{a}\star
_{K}x^{c}\right) \\
&=&M_{b}^{a}x^{b}
\end{eqnarray*}
with 
\begin{equation*}
M_{b}^{a}=-\frac{1}{2}\theta ^{ac}\Delta _{cb}.
\end{equation*}
It is then clear that 
\begin{eqnarray*}
C(x) &=&x^{1}\star _{\omega }\cdots \star _{\omega }x^{n-2}\;\pi _{1\cdots
n-2} \\
\pi _{1\cdots n-2} &=&M_{1}^{a}\tau _{a2\cdots n-2}+\cdots +M_{n-2}^{a}\tau
_{1\cdots n-3,a}
\end{eqnarray*}
It is now quite easy to show that $\pi $ is cyclic gauge invariant,
therefore implying that $C=0$. $\square $

\begin{lemma}
\label{L-dbe}(Section \ref{giiuact}) Let $\eta _{1\cdots n+2}$ be gauge
invariant, and let us consider the combination $C=\left( \lambda ^{1}\star
\cdots \star \lambda ^{n+2}\right) \,\eta _{1\cdots n+2}$. If we add a
central term $\Delta ^{ab}$ to the commutator $-i[\lambda ^{a},\lambda ^{b}]$%
, then the expression $C$ varies by $-\left( \lambda ^{1}\star \cdots \star
\lambda ^{n}\right) \left( \overline{\Delta }\eta \right) _{1\cdots n}$,
where 
\begin{equation*}
\left( \overline{\Delta }\eta \right) ^{1\cdots n}=-\frac{i}{2}\Delta
^{ab}\left( \eta _{1\cdots nab}+\eta _{1\cdots anb}+\cdots \right) .
\end{equation*}
Moreover, for any $\Delta ^{ab}$, the tensor $\overline{\Delta }\eta $ is
gauge invariant.
\end{lemma}

P{\small ROOF}. First it is clear that, if $\eta _{1\cdots n}$ and $\nu
_{1\cdots m}$ are gauge invariant, then so is $(\eta \nu )_{1\cdots n+m}=$ $%
\eta _{1\cdots n}\nu _{n+1\cdots n+m}$. Moreover, if $d_{a}$ is any
one--indexed tensor, then $(d\eta )_{1\cdots n+1}=d_{1}\eta _{2\cdots
n+1}-\eta _{1\cdots n}d_{n+1}$ is again gauge invariant. These two facts can
either be checked algebraically, or one can simply note that they correspond
respectively to the product and covariant derivative of gauge invariant
operators. In fact, any gauge invariant tensor is built using the two
operations just described.

In order to prove the lemma we first show that it holds for $n=0$. Then $%
\eta _{ab}$ is an antisymmetric tensor and $\lambda ^{a}\star \lambda
^{b}\,\eta _{ab}=\frac{1}{2}[\lambda ^{a},\lambda ^{b}]\,\eta
_{ab}\rightarrow \frac{1}{2}[\lambda ^{a},\lambda ^{b}]\,\eta _{ab}+\frac{i}{%
2}\Delta ^{ab}\eta _{ab}$. Now suppose that we have proved the result for $%
\eta _{1\cdots n}$ and $\nu _{1\cdots m}$. Then we must show that $\overline{%
\Delta }(\eta \nu )=(\overline{\Delta }\eta )\nu +\eta (\overline{\Delta }%
\nu )$. This is easily done, since 
\begin{eqnarray*}
\overline{\Delta }(\eta \nu )-(\overline{\Delta }\eta )\nu -\eta (\overline{%
\Delta }\nu ) &\varpropto &\Delta ^{ab}(\eta _{a,1,\cdots ,n-1}+\eta
_{1,a,\cdots ,n-1}+\cdots +\eta _{1,\cdots ,n-1,a})\cdot \\
&&\cdot (\nu _{b,n\cdots ,n+m-2}+\nu _{n,b,\cdots ,n+m-2}+\cdots +\nu
_{n,\cdots ,n+m-2,b})
\end{eqnarray*}
which vanishes since $\eta $ and $\nu $ are gauge invariant. Finally me must
show that, given a generic $d_{a}$, one has $\overline{\Delta }(d\eta )=d(%
\overline{\Delta }\eta )$. Again this is easy to show using the gauge
invariance of $\eta $, since 
\begin{equation*}
\overline{\Delta }(d\eta )-d(\overline{\Delta }\eta )\varpropto \Delta
^{ab}d_{a}(\eta _{b,1,\cdots ,n}+\eta _{1,b,\cdots ,n}+\cdots +\eta
_{1,\cdots ,n,b})=0.
\end{equation*}
This concludes the proof, since any gauge invariant operator is a product of
covariant derivatives of the field strength. $\square $

\begin{lemma}
\label{L-3}(Section \ref{bict}) Let $\eta _{1\cdots n+2}$ be gauge invariant
and let $\tau _{1\cdots n+2}$ be the associated cyclic gauge invariant
tensor. Let $\Delta ^{ab}$ be antisymmetric, and define 
\begin{equation}
g_{1\cdots n}=\frac{1}{n}\left( \overline{\Delta }\eta _{1\cdots n}+\mathrm{%
cyc}_{1\cdots n}\right)  \label{A2000}
\end{equation}
where $\overline{\Delta }\eta $ is given by expression (\ref{eq2000}). The
tensor $g$ is then uniquely a function of $\tau $, and is explicitly given
by the expression 
\begin{equation}
-\frac{i}{2}\left( \frac{n+2}{n^{2}}\right) \Delta ^{ab}\left[ n\tau
_{1\cdots nab}+\left( n-1\right) \tau _{1\cdots anb}+\cdots +0\cdot \tau
_{a1\cdots nb}\right] +\mathrm{cyc}_{1\cdots n}  \label{A3000}
\end{equation}
which will be denoted by $\overline{\Delta }\tau $.
\end{lemma}

P{\small ROOF}. Define the following tensors 
\begin{eqnarray*}
k_{1\cdots nab} &=&\eta _{1\cdots nab}+\mathrm{cyc}_{1\cdots n} \\
k_{1\cdots anb} &=&\eta _{1\cdots anb}+\mathrm{cyc}_{1\cdots n} \\
k_{1\cdots n-1,b,a,n} &=&\eta _{1\cdots n-1,b,a,n}+\mathrm{cyc}_{1\cdots n}
\\
&&\cdots
\end{eqnarray*}
which have two selected indices $a,b$ and are cyclically symmetric in the
other indices $1,\cdots ,n$. It is clear that 
\begin{eqnarray*}
\tau _{1\cdots nab}+\mathrm{cyc}_{1\cdots n} &=&\frac{1}{n+2}\left(
k_{1\cdots nab}+\mathrm{cyc}_{1\cdots nab}\right) \\
\tau _{1\cdots anb}+\mathrm{cyc}_{1\cdots n} &=&\frac{1}{n+2}\left(
k_{1\cdots anb}+\mathrm{cyc}_{1\cdots anb}\right) \\
&&\dots
\end{eqnarray*}
We can also express equation (\ref{A2000}) for $g$ in terms of the tensors $%
k $ as 
\begin{eqnarray*}
g_{1\cdots n} &=&-\frac{i}{2n^{2}}\Delta ^{ab}J_{1\cdots nab} \\
J_{1\cdots nab} &=&nk_{1\cdots nab}+nk_{1\cdots anb}+\cdots ,
\end{eqnarray*}
where, as in $\overline{\Delta }\eta $, the indices $a,b$ are in all
possible positions with $a$ preceding $b$. We then want to prove that the
above expression is equal to (\ref{A3000}), which we can also write in terms
of the tensors $k$ as follows 
\begin{equation*}
-\frac{i}{2n^{2}}\Delta ^{ab}\widetilde{J}_{1\cdots nab}
\end{equation*}
with 
\begin{eqnarray*}
\widetilde{J}_{1\cdots nab} &=&n\left( k_{1\cdots nab}+\mathrm{cyc}_{1\cdots
nab}\right) \\
&&+\left( n-1\right) \left( k_{1\cdots anb}+\mathrm{cyc}_{1\cdots anb}\right)
\\
&&+\cdots \\
&&+0\cdot \left( k_{a1\cdots nb}+\mathrm{cyc}_{a1\cdots nb}\right)
\end{eqnarray*}
We will actually prove that $J_{1\cdots nab}=\widetilde{J}_{1\cdots nab}$.
In order to do this, we use the gauge invariance of the tensor $\eta $,
which implies various linear relations among the tensors $k$ 
\begin{eqnarray}
0 &=&k_{ab1\cdots n}+k_{ba1\cdots n}+k_{b1a\cdots n}+\cdots +k_{b1\cdots na}
\label{A10000} \\
0 &=&k_{a1b\cdots n}+k_{1ab\cdots n}+k_{1ba\cdots n}+\cdots +k_{1b2\cdots na}
\notag \\
&&\cdots  \notag
\end{eqnarray}
We then need to prove that the difference $J-\widetilde{J}$ can be written
as a linear combination of the above equations.

In order to write an efficient and clear proof, we will concentrate, from
now on, on the case $n=4$. The proof in the general case is absolutely
identical, but the added notation would obscure the result without adding
new ideas to the ones already contained in the special case discussed below.
We introduce the following compact notation 
\begin{eqnarray*}
k_{1234ab} &\rightarrow & 
\begin{array}{cccccc}
{\small \cdot } & {\small \cdot } & {\small \cdot } & {\small \cdot } & 
{\small a} & {\small b}
\end{array}
\\
k_{123a4b} &\rightarrow & 
\begin{array}{cccccc}
{\small \cdot } & {\small \cdot } & {\small \cdot } & {\small a} & {\small %
\cdot } & {\small b}
\end{array}
\\
&&\cdots
\end{eqnarray*}
where the dots $
\begin{array}{cccc}
\cdot & \cdot & \cdot & \cdot
\end{array}
$ indicate the indices $1,2,3,4$. We then arrange all the possible tensors $%
k $ in the following tableau 
\begin{equation*}
\begin{tabular}{|c|c|}
\hline
$
\begin{array}{cccccc}
\cdot & \cdot & \cdot & \cdot & a & b \\ 
\cdot & \cdot & \cdot & a & b & \cdot \\ 
\cdot & \cdot & a & b & \cdot & \cdot \\ 
\cdot & a & b & \cdot & \cdot & \cdot \\ 
a & b & \cdot & \cdot & \cdot & \cdot
\end{array}
$ & $
\begin{array}{cccccc}
\cdot & \cdot & \cdot & \cdot & b & a \\ 
\cdot & \cdot & \cdot & b & a & \cdot \\ 
\cdot & \cdot & b & a & \cdot & \cdot \\ 
\cdot & b & a & \cdot & \cdot & \cdot \\ 
b & a & \cdot & \cdot & \cdot & \cdot
\end{array}
$ \\ \hline
$
\begin{array}{cccccc}
\cdot & \cdot & \cdot & a & \cdot & b \\ 
\cdot & \cdot & a & \cdot & b & \cdot \\ 
\cdot & a & \cdot & b & \cdot & \cdot \\ 
a & \cdot & b & \cdot & \cdot & \cdot
\end{array}
$ & $
\begin{array}{cccccc}
\cdot & \cdot & \cdot & b & \cdot & a \\ 
\cdot & \cdot & b & \cdot & a & \cdot \\ 
\cdot & b & \cdot & a & \cdot & \cdot \\ 
b & \cdot & a & \cdot & \cdot & \cdot
\end{array}
$ \\ \hline
$
\begin{array}{cccccc}
\cdot & \cdot & a & \cdot & \cdot & b \\ 
\cdot & a & \cdot & \cdot & b & \cdot \\ 
a & \cdot & \cdot & b & \cdot & \cdot
\end{array}
$ & $
\begin{array}{cccccc}
\cdot & \cdot & b & \cdot & \cdot & a \\ 
\cdot & b & \cdot & \cdot & a & \cdot \\ 
b & \cdot & \cdot & a & \cdot & \cdot
\end{array}
$ \\ \hline
$
\begin{array}{cccccc}
\cdot & a & \cdot & \cdot & \cdot & b \\ 
a & \cdot & \cdot & \cdot & b & \cdot
\end{array}
$ & $
\begin{array}{cccccc}
\cdot & b & \cdot & \cdot & \cdot & a \\ 
b & \cdot & \cdot & \cdot & a & \cdot
\end{array}
$ \\ \hline
$
\begin{array}{cccccc}
a & \cdot & \cdot & \cdot & \cdot & b
\end{array}
$ & $
\begin{array}{cccccc}
b & \cdot & \cdot & \cdot & \cdot & a
\end{array}
$ \\ \hline
\end{tabular}
\end{equation*}
which has on the left all terms with $a$ preceding $b$, and on the right all
terms with $b$ before $a$. From top to bottom, the terms are, on the other
hand, arranged in groups with a fixed number of indices between $a$ and $b$.
We then denote any linear combination of the tensors $k$ with a horizontal
box of coefficients 
\begin{equation*}
\begin{tabular}{|c|c|c|c|c|}
\hline
$a_{1}a_{2}a_{3}a_{4}a_{5}$ & $b_{1}b_{2}b_{3}b_{4}$ & $c_{1}c_{2}c_{3}$ & $%
d_{1}d_{2}$ & $e_{1}$ \\ \hline
$f_{1}f_{2}f_{3}f_{4}f_{5}$ & $g_{1}g_{2}g_{3}g_{4}$ & $h_{1}h_{2}h_{3}$ & $%
i_{1}i_{2}$ & $j_{1}$ \\ \hline
\end{tabular}
\end{equation*}
where the top line corresponds to the coefficients of the left column in the
tableau, and the bottom line to the right column (the above box is then a
compact notation for the sum $a_{1}k_{1234ab}+a_{2}k_{123a4b}+\cdots
+f_{1}k_{1234ba}+\cdots $). The main statement which we want to prove can
now be compactly written as 
\begin{equation}
\begin{tabular}{|c|c|c|c|c|}
\hline
$44444$ & $4444$ & $444$ & $44$ & $4$ \\ \hline
$00000$ & $0000$ & $000$ & $00$ & $0$ \\ \hline
\end{tabular}
= 
\begin{tabular}{|c|c|c|c|c|}
\hline
$44444$ & $3333$ & $222$ & $11$ & $0$ \\ \hline
$00000$ & $1111$ & $222$ & $33$ & $4$ \\ \hline
\end{tabular}
\label{A10001}
\end{equation}
In fact the LHS above is nothing but $J$. To show that the RHS is $%
\widetilde{J}$ we just note that 
\begin{eqnarray*}
k_{1\cdots nab}+\mathrm{cyc}_{1\cdots nab} &=& 
\begin{tabular}{|c|c|c|c|c|}
\hline
$11111$ & $0000$ & $000$ & $00$ & $0$ \\ \hline
$00000$ & $0000$ & $000$ & $00$ & $1$ \\ \hline
\end{tabular}
\\
k_{1\cdots anb}+\mathrm{cyc}_{1\cdots anb} &=& 
\begin{tabular}{|c|c|c|c|c|}
\hline
$00000$ & $1111$ & $000$ & $00$ & $0$ \\ \hline
$00000$ & $0000$ & $000$ & $11$ & $0$ \\ \hline
\end{tabular}
\\
&&\dots
\end{eqnarray*}
To prove the equality (\ref{A10001}) we use the linear relations (\ref
{A10000}), which can also be compactly rewritten as 
\begin{equation*}
a_{i}=b_{j}=0,
\end{equation*}
where 
\begin{eqnarray*}
a_{1} &=& 
\begin{tabular}{|c|c|c|c|c|}
\hline
$10000$ & $1000$ & $100$ & $10$ & $1$ \\ \hline
$10000$ & $0000$ & $000$ & $00$ & $0$ \\ \hline
\end{tabular}
\,\ \ \ \ \ \ \ b_{1}= 
\begin{tabular}{|c|c|c|c|c|}
\hline
$00001$ & $0001$ & $001$ & $01$ & $1$ \\ \hline
$00001$ & $0000$ & $000$ & $00$ & $0$ \\ \hline
\end{tabular}
\\
a_{2} &=& 
\begin{tabular}{|c|c|c|c|c|}
\hline
$01000$ & $0100$ & $010$ & $01$ & $0$ \\ \hline
$01000$ & $1000$ & $000$ & $00$ & $0$ \\ \hline
\end{tabular}
\,\ \ \ \ \ \ \ b_{2}= 
\begin{tabular}{|c|c|c|c|c|}
\hline
$00010$ & $0010$ & $010$ & $10$ & $0$ \\ \hline
$00010$ & $0001$ & $000$ & $00$ & $0$ \\ \hline
\end{tabular}
\\
a_{3} &=& 
\begin{tabular}{|c|c|c|c|c|}
\hline
$00100$ & $0010$ & $001$ & $00$ & $0$ \\ \hline
$00100$ & $0100$ & $100$ & $00$ & $0$ \\ \hline
\end{tabular}
\,\ \ \ \ \ \ \ b_{3}= 
\begin{tabular}{|c|c|c|c|c|}
\hline
$00100$ & $0100$ & $100$ & $00$ & $0$ \\ \hline
$00100$ & $0010$ & $001$ & $00$ & $0$ \\ \hline
\end{tabular}
\\
a_{4} &=& 
\begin{tabular}{|c|c|c|c|c|}
\hline
$00010$ & $0001$ & $000$ & $00$ & $0$ \\ \hline
$00010$ & $0010$ & $010$ & $10$ & $0$ \\ \hline
\end{tabular}
\,\ \ \ \ \ \ \ b_{4}= 
\begin{tabular}{|c|c|c|c|c|}
\hline
$01000$ & $1000$ & $000$ & $00$ & $0$ \\ \hline
$01000$ & $0100$ & $010$ & $01$ & $0$ \\ \hline
\end{tabular}
\\
a_{5} &=& 
\begin{tabular}{|c|c|c|c|c|}
\hline
$00001$ & $0000$ & $000$ & $00$ & $0$ \\ \hline
$00001$ & $0001$ & $001$ & $01$ & $1$ \\ \hline
\end{tabular}
\,\ \ \ \ \ \ \ b_{5}= 
\begin{tabular}{|c|c|c|c|c|}
\hline
$10000$ & $0000$ & $000$ & $00$ & $0$ \\ \hline
$10000$ & $1000$ & $100$ & $10$ & $1$ \\ \hline
\end{tabular}
\end{eqnarray*}
Summing either all the $a$'s or all the $b$'s we first of all obtain the
following interesting identity 
\begin{equation}
\begin{tabular}{|c|c|c|c|c|}
\hline
$11111$ & $1111$ & $111$ & $11$ & $1$ \\ \hline
$11111$ & $1111$ & $111$ & $11$ & $1$ \\ \hline
\end{tabular}
=0.  \label{AAA321}
\end{equation}
Combining the above equation with (\ref{A10001}) we can then reduce the
statement of the lemma to the following equality 
\begin{equation*}
\begin{tabular}{|c|c|c|c|c|}
\hline
$44444$ & $5555$ & $666$ & $77$ & $8$ \\ \hline
$44444$ & $3333$ & $222$ & $11$ & $0$ \\ \hline
\end{tabular}
=0.
\end{equation*}
The LHS above is nothing but 
\begin{equation*}
4(a_{1}+b_{1})+3(a_{2}+b_{2})+2(a_{3}+b_{3})+1(a_{4}+b_{4})+0(a_{5}+b_{5}),
\end{equation*}
and therefore the proof is complete. $\square $

\begin{remark}
Using the box (\ref{AAA321}) we can show that the expression 
\begin{equation*}
\left[ n\tau _{1\cdots nab}+\left( n-1\right) \tau _{1\cdots anb}+\cdots
+0\cdot \tau _{a1\cdots nb}\right] +\mathrm{cyc}_{1\cdots n}
\end{equation*}
in (\ref{A3000}) is antisymmetric in $a$ and $b$.
\end{remark}

\begin{lemma}
(Section \ref{iuctbc}) \label{L-7}Let $\tau _{1\cdots n}$ be a cyclic gauge
invariant tensor, and let $\Delta _{ab}$ be an arbitrary invertible
antisymmetric matrix with inverse $\Delta ^{ab}$. Then 
\begin{equation*}
\lbrack \overline{\Delta },\Delta ]\tau =\left( n-\frac{1}{2}D\right) \tau .
\end{equation*}
\end{lemma}

P{\small ROOF}. First we recall the expression for $\Delta \tau $, which is
given by 
\begin{equation*}
\left( \Delta \tau \right) _{1\cdots n+2}=\frac{i}{2}\left( \frac{n}{n+2}%
\right) \left( \Delta _{12}\tau _{345\cdots n+2}-\Delta _{13}\tau
_{245\cdots n+2}\right) +\mathrm{cyc}_{1\cdots n+2}.
\end{equation*}
We then concentrate on the expression $\overline{\Delta }\Delta \tau $ by
recalling, first of all, that the operation $\overline{\Delta }$ on $\Delta
\tau $ consists of a contraction of two indices of the tensor $\Delta \tau $
with the antisymmetric tensor $\Delta ^{ab}$. It is then clear, since $%
\Delta \tau $ itself is built from the tensors $\tau _{1\cdots n}$ and $%
\Delta _{ab}$, that 
\begin{equation*}
\overline{\Delta }\Delta \tau =A+B+C,
\end{equation*}
where $A$ contains the terms where $\Delta ^{ab}$ is contracted uniquely
with $\Delta _{ab}$ and $B$ contains terms in which the two indices in $%
\Delta ^{ab}$ are contracted one with $\Delta _{ab}$ and one with $\tau
_{1\cdots n}$. Finally $C$ consists of the remaining terms, with
contractions of $\Delta ^{ab}$ only with $\tau _{1\cdots n}$. We will prove
in the sequel that 
\begin{eqnarray*}
A &=&-\frac{1}{2}D\tau \\
B &=&n\tau \\
C &=&\Delta \overline{\Delta }\tau ,
\end{eqnarray*}
thus proving the lemma.

Let me start by concentrating on the terms in $A$. First recall the
expression for $\overline{\Delta }\Delta \tau $%
\begin{equation}
\overline{\Delta }\Delta \tau =-\frac{i}{2}\left( \frac{n+2}{n^{2}}\right)
\Delta ^{ab}\left[ n\cdot \left( \Delta \tau \right) _{1\cdots nab}+\cdots
+1\cdot \left( \Delta \tau \right) _{1a2\cdots nb}\right] +\mathrm{cyc}%
_{1\cdots n}  \label{A111}
\end{equation}
The only terms in the above equation which contribute to $A$ are the first,
second and last within the square bracket. In particular the first term
reads 
\begin{eqnarray*}
&&-\frac{i}{2}\left( \frac{n+2}{n}\right) \Delta ^{ab}\left( \Delta \tau
\right) _{1\cdots nab}+\mathrm{cyc}_{1\cdots n} \\
&=&\frac{1}{4}\Delta ^{ab}\left[ \left( \Delta _{12}\tau _{3\cdots
nab}-\Delta _{13}\tau _{24\cdots nab}\right) +\mathrm{cyc}_{1\cdots nab}%
\right] +\mathrm{cyc}_{1\cdots n} \\
&=&\frac{1}{4}\Delta ^{ab}\Delta _{ab}\tau _{1\cdots n}+\mathrm{cyc}%
_{1\cdots n}+\mathrm{terms\;not\;in\;}A \\
&=&-n\frac{D}{4}\tau _{1\cdots n}+\mathrm{terms\;not\;in\;}A,
\end{eqnarray*}
where we have used that 
\begin{equation*}
\Delta _{ab}\Delta ^{ab}=-D
\end{equation*}
and that $\tau $ is cyclic. Similarly, the second term is given by 
\begin{eqnarray*}
&&-\frac{i}{2}\left( \frac{n+2}{n^{2}}\right) (n-1)\Delta ^{ab}\left( \Delta
\tau \right) _{1\cdots anb}+\mathrm{cyc}_{1\cdots n} \\
&=&\left( \frac{n-1}{4n}\right) \Delta ^{ab}\left[ \left( \Delta _{12}\tau
_{3\cdots anb}-\Delta _{13}\tau _{24\cdots anb}\right) +\mathrm{cyc}%
_{1\cdots anb}\right] +\mathrm{cyc}_{1\cdots n} \\
&=&\left( n-1\right) \frac{D}{4}\tau _{1\cdots n}+\mathrm{terms\;not\;in\;}A.
\end{eqnarray*}
Finally the last term 
\begin{eqnarray*}
&&-\frac{i}{2}\left( \frac{n+2}{n^{2}}\right) \Delta ^{ab}\left( \Delta \tau
\right) _{1a2\cdots nb}+\mathrm{cyc}_{1\cdots n} \\
&=&\frac{1}{4n}\Delta ^{ab}\left[ \left( \Delta _{1a}\tau _{2\cdots
nb}-\Delta _{12}\tau _{a3\cdots nb}\right) +\mathrm{cyc}_{1a2\cdots nb}%
\right] +\mathrm{cyc}_{1\cdots n} \\
&=&-\frac{D}{4}\tau _{1\cdots n}+\mathrm{terms\;not\;in\;}A
\end{eqnarray*}
Summing the three contributions we obtain 
\begin{equation*}
A=-\frac{D}{2}\tau
\end{equation*}
as was to be shown.

We now move to the analysis of the terms in $B$. Again we consider equation (%
\ref{A111}), and we focus once again on the first term in the square
brackets 
\begin{equation*}
\frac{1}{4}\Delta ^{ab}\left[ \left( \Delta _{12}\tau _{3\cdots nab}-\Delta
_{13}\tau _{24\cdots nab}\right) +\mathrm{cyc}_{1\cdots nab}\right] +\mathrm{%
cyc}_{1\cdots n}
\end{equation*}
We concentrate on the terms in $B$, with $\Delta ^{ab}$ contracted both with 
$\Delta _{ab}$ and with $\tau _{1\cdots n}$ 
\begin{eqnarray}
&&\frac{1}{4}\Delta ^{ab}\left[ \Delta _{na}\tau _{b1\cdots n-1}+\Delta
_{b1}\tau _{2\cdots na}\right] +\mathrm{cyc}_{1\cdots n}  \notag \\
&&-\frac{1}{4}\Delta ^{ab}\left[ \Delta _{a1}\tau _{b2\cdots n}+\Delta
_{nb}\tau _{a1\cdots n-1}+\Delta _{n-1,a}\tau _{nb1\cdots n-2}+\Delta
_{b2}\tau _{13\cdots na}\right] +\mathrm{cyc}_{1\cdots n}  \notag \\
&=&\frac{1}{4n}\left( 4n\,\tau _{1\cdots n}-n\,\tau _{213\cdots n}-n\,\tau
_{23\cdots 1n}\right) +\mathrm{cyc}_{1\cdots n}  \label{A333}
\end{eqnarray}
The second term in (\ref{A111}) 
\begin{equation*}
\left( \frac{n-1}{4n}\right) \Delta ^{ab}\left[ \left( \Delta _{12}\tau
_{3\cdots anb}-\Delta _{13}\tau _{24\cdots anb}\right) +\mathrm{cyc}%
_{1\cdots anb}\right] +\mathrm{cyc}_{1\cdots n}
\end{equation*}
gives as contribution to $B$%
\begin{eqnarray}
&&\left( \frac{n-1}{4n}\right) \Delta ^{ab}\left[ \Delta _{b1}\tau
_{23\cdots an}+\Delta _{nb}\tau _{1\cdots n-1a}+\Delta _{n-1,a}\tau
_{nb1\cdots n-2}+\Delta _{an}\tau _{b1\cdots n-1}\right] +\mathrm{cyc}%
_{1\cdots n}  \notag \\
&&-\left( \frac{n-1}{4n}\right) \Delta ^{ab}\left[ \Delta _{b2}\tau
_{13\cdots an}+\Delta _{n-2,a}\tau _{n-1,n,b,1\cdots ,n-3}\right] +\mathrm{%
cyc}_{1\cdots n}  \notag \\
&=&\left( \frac{n-1}{4n}\right) \left[ \tau _{213\cdots n}+\tau _{23\cdots
1n}-\tau _{231\cdots n}-\tau _{23\cdots ,1,n-1,n}-2\tau _{1\cdots n}\right] +%
\mathrm{cyc}_{1\cdots n}  \label{A444}
\end{eqnarray}
In order to write equations in a compact form let us introduce some
notation. We explain the notation in the case $n=4$, but then we continue
the proof in a general setting. We wish to consider tensors $\tau _{\cdots }$
with the indices given by $2,3,4$ in increasing order, and with the index $1$
in a given position. For example, if the index $1$ is in the $3$rd position
we are considering the tensor $\tau _{2314}$, which we will denote with the
following box 
\begin{equation*}
\tau _{2314}\rightarrow 
\begin{tabular}{|c|c|c|c|}
\hline
$0$ & $0$ & $1$ & $0$ \\ \hline
\end{tabular}
\,.
\end{equation*}
Moreover, linear combinations of the various tensors will also be denoted by
a single box in the following obvious way 
\begin{eqnarray*}
\begin{tabular}{|c|c|c|c|}
\hline
$a$ & $b$ & $c$ & $d$ \\ \hline
\end{tabular}
\ &=&a\, 
\begin{tabular}{|c|c|c|c|}
\hline
$1$ & $0$ & $0$ & $0$ \\ \hline
\end{tabular}
+b\, 
\begin{tabular}{|c|c|c|c|}
\hline
$0$ & $1$ & $0$ & $0$ \\ \hline
\end{tabular}
\\
&&+c\, 
\begin{tabular}{|c|c|c|c|}
\hline
$0$ & $0$ & $1$ & $0$ \\ \hline
\end{tabular}
+d\, 
\begin{tabular}{|c|c|c|c|}
\hline
$0$ & $0$ & $0$ & $1$ \\ \hline
\end{tabular}
\,.
\end{eqnarray*}
Let us now return to the general proof, by first showing some simple
properties of the box just introduced. Cyclicity of the tensor $\tau $
implies that 
\begin{equation}
\begin{tabular}{|c|c|c|c|c|}
\hline
$1$ & $0$ & $\cdots $ & $0$ & $0$ \\ \hline
\end{tabular}
= 
\begin{tabular}{|c|c|c|c|c|}
\hline
$0$ & $0$ & $\cdots $ & $0$ & $1$ \\ \hline
\end{tabular}
\,.  \label{A555}
\end{equation}
Moreover, cyclic gauge invariance of $\tau $ implies the two equivalent
identities 
\begin{equation*}
\begin{tabular}{|c|c|c|c|c|}
\hline
$1$ & $1$ & $\cdots $ & $1$ & $0$ \\ \hline
\end{tabular}
= 
\begin{tabular}{|c|c|c|c|c|}
\hline
$0$ & $1$ & $\cdots $ & $1$ & $1$ \\ \hline
\end{tabular}
=0
\end{equation*}
which can be summed to obtain 
\begin{equation}
\begin{tabular}{|c|c|c|c|c|}
\hline
$1$ & $2$ & $\cdots $ & $2$ & $1$ \\ \hline
\end{tabular}
=0.  \label{A222}
\end{equation}
We can use this notation to compactly rewrite the first two terms in $%
\overline{\Delta }\Delta \tau $ given by equations (\ref{A333}) and (\ref
{A444}). They now read 
\begin{equation*}
\frac{1}{4n}\; 
\begin{tabular}{|c|c|c|c|c|c|c|}
\hline
$2n$ & $-n$ & $0$ & $\cdots $ & $0$ & $-n$ & $2n$ \\ \hline
\end{tabular}
+\mathrm{cyc}_{1\cdots n}
\end{equation*}
and 
\begin{equation*}
\frac{1}{4n}\; 
\begin{tabular}{|c|c|c|c|c|c|c|}
\hline
$-(n-1)$ & $n-1$ & $-(n-1)$ & $\cdots $ & $-(n-1)$ & $n-1$ & $-(n-1)$ \\ 
\hline
\end{tabular}
+\mathrm{cyc}_{1\cdots n}
\end{equation*}
We notice that both terms are represented by boxes which are symmetric about
a vertical axis of symmetry. Moreover we can do computations similar to the
ones above to convince ourselves that the $(k+1)$--th term in $\overline{%
\Delta }\Delta \tau $ is given by the following box (we just show the left
side of the box, since the right side is just its mirror copy) 
\begin{equation*}
\frac{1}{4n}\; 
\begin{tabular}{|c|c|c|c|c|c|c|c}
\hline
$0$ & $\cdots $ & $0$ & $n-k$ & $-(n-k)$ & $n-k$ & $-(n-k)$ & $\cdots $ \\ 
\hline
\end{tabular}
+\mathrm{cyc}_{1\cdots n}
\end{equation*}
with $k-2$ zeros on the left before the term $n-k$. Writing at once all the
contributions to $B$ we get the following tableau 
\begin{equation*}
\frac{1}{4n}\; 
\begin{tabular}{|c|c|c|c|c}
\hline
$2n$ & $-n$ &  &  &  \\ \hline
$-(n-1)$ & $n-1$ & $-(n-1)$ &  &  \\ \hline
$n-2$ & $-(n-2)$ & $n-2$ & $-(n-2)$ &  \\ \hline
& $n-3$ & $-(n-3)$ & $n-3$ & $-(n-3)$ \\ \hline
&  &  &  & 
\end{tabular}
+\mathrm{cyc}_{1\cdots n}
\end{equation*}
where we must sum the coefficients in each column. The result is then 
\begin{eqnarray*}
&&\frac{1}{4n}\, 
\begin{tabular}{|c|c|c|c|c|c|c|}
\hline
$2n-1$ & $-2$ & $-2$ & $\cdots $ & $-2$ & $-2$ & $2n-1$ \\ \hline
\end{tabular}
+\mathrm{cyc}_{1\cdots n} \\
&=&\frac{1}{2}\, 
\begin{tabular}{|c|c|c|c|c|}
\hline
$1$ & $0$ & $\cdots $ & $0$ & $1$ \\ \hline
\end{tabular}
+\mathrm{cyc}_{1\cdots n}= 
\begin{tabular}{|c|c|c|c|}
\hline
$1$ & $0$ & $\cdots $ & $0$ \\ \hline
\end{tabular}
+\mathrm{cyc}_{1\cdots n}
\end{eqnarray*}
where we have used equations (\ref{A222}) and (\ref{A555}). Going back to
the usual tensor notation we have then obtained 
\begin{equation*}
B=\tau _{1\cdots n}+\mathrm{cyc}_{1\cdots n}=n\tau .
\end{equation*}

We conclude the proof by showing that $C=\Delta \overline{\Delta }\tau $.
Consider again the first term in $\overline{\Delta }\Delta \tau $%
\begin{equation*}
\frac{1}{4}\Delta ^{ab}\left[ \left( \Delta _{12}\tau _{3\cdots nab}-\Delta
_{13}\tau _{24\cdots nab}\right) +\mathrm{cyc}_{1\cdots nab}\right] +\mathrm{%
cyc}_{1\cdots n}
\end{equation*}
and concentrate on the terms which contain $\tau _{\cdots ab}$ -- \textit{%
i.e. }terms for which the indices in $\tau $ contain $a$ just before $b$ -- 
\begin{eqnarray}
&&\frac{1}{4}\left( \Delta _{12}\Delta ^{ab}\left( \tau _{3\cdots nab}+%
\mathrm{cyc}_{3\cdots n}\right) +\mathrm{cyc}_{1\cdots n}\right)
\label{eqPPP1} \\
&&-\frac{1}{4}\left( \Delta _{13}\Delta ^{ab}\left( \tau _{24\cdots nab}+%
\mathrm{cyc}_{24\cdots n}\right) +\mathrm{cyc}_{1\cdots n}\right)  \notag \\
&&+\frac{1}{4}\left( \Delta _{12}\Delta ^{ab}\tau _{3\cdots nab}+\mathrm{cyc}%
_{1\cdots n}\right)  \notag
\end{eqnarray}
Terms with $\tau _{\cdots ab}$ are also contained in the second and third
contribution to $\Delta \overline{\Delta }\tau $ in (\ref{A111}) 
\begin{eqnarray*}
&&\left( \frac{n-1}{4n}\right) \Delta ^{ab}\left[ \left( \Delta _{12}\tau
_{3\cdots anb}-\Delta _{13}\tau _{24\cdots anb}\right) +\mathrm{cyc}%
_{1\cdots anb}\right] +\mathrm{cyc}_{1\cdots n} \\
&&\left( \frac{n-2}{4n}\right) \Delta ^{ab}\left[ \left( \Delta _{12}\tau
_{3\cdots a,n-1,nb}-\Delta _{13}\tau _{24\cdots a,n-1,nb}\right) +\mathrm{cyc%
}_{1\cdots a,n-1,n,b}\right] +\mathrm{cyc}_{1\cdots n}
\end{eqnarray*}
They are respectively 
\begin{equation}
-2\left( \frac{n-1}{4n}\right) \left( \Delta _{12}\Delta ^{ab}\tau _{3\cdots
nab}+\mathrm{cyc}_{1\cdots n}\right)  \label{eqPPP2}
\end{equation}
and 
\begin{equation}
\left( \frac{n-2}{4n}\right) \left( \Delta _{12}\Delta ^{ab}\tau _{3\cdots
nab}+\mathrm{cyc}_{1\cdots n}\right) .  \label{eqPPP3}
\end{equation}
Summing equations (\ref{eqPPP1}),(\ref{eqPPP2}) and (\ref{eqPPP3}) we then
obtain (the last line in (\ref{eqPPP1}) is canceled by (\ref{eqPPP2}) and (%
\ref{eqPPP3})) 
\begin{eqnarray*}
&&\frac{1}{4}\left( \Delta _{12}\Delta ^{ab}\left( \tau _{3\cdots nab}+%
\mathrm{cyc}_{3\cdots n}\right) +\mathrm{cyc}_{1\cdots n}\right) \\
&&-\frac{1}{4}\left( \Delta _{13}\Delta ^{ab}\left( \tau _{24\cdots nab}+%
\mathrm{cyc}_{24\cdots n}\right) +\mathrm{cyc}_{1\cdots n}\right) .
\end{eqnarray*}
But this is exactly the result which is obtained by computing terms
proportional to $\tau _{\cdots ab}$ in $\Delta \overline{\Delta }\tau $,
since 
\begin{equation*}
\Delta \overline{\Delta }\tau =\frac{i}{2}\left( \frac{n-2}{n}\right) \left[
\Delta _{12}\left( \overline{\Delta }\tau \right) _{34\cdots n}-\Delta
_{13}\left( \overline{\Delta }\tau \right) _{24\cdots n}+\mathrm{cyc}%
_{1\cdots n}\right]
\end{equation*}
and, if we concentrate on terms of the form $\tau _{\cdots ab}$, 
\begin{equation*}
\left( \overline{\Delta }\tau \right) _{34\cdots n}=-\frac{i}{2}\left( \frac{%
n}{n-2}\right) \Delta ^{ab}\left( \tau _{3\cdots nab}+\mathrm{cyc}_{3\cdots
n}\right) .
\end{equation*}
Similar arguments can be used for terms proportional to $\tau _{\cdots
a\cdots b}$, with different number of indices between $a$ and $b$. We have
then shown that $C=\Delta \overline{\Delta }\tau $, thus completing the
proof. $\square $

\begin{lemma}
\label{L-10}(Section \ref{ie2d}) If $F=\left[ Z,\overline{Z}\right] $, then 
\begin{equation*}
\Delta F^{3}=2F^{4}+\frac{1}{2}F[DF,\overline{D}F].
\end{equation*}
\end{lemma}

P{\small ROOF}. We compute the word $w\in W_{4}$ 
\begin{equation*}
w=\Delta F^{3}.
\end{equation*}
First we have that 
\begin{equation*}
w=3F^{2}\left( \Delta F\right) =\frac{3}{2}\left( F^{4}+F\left[ FZ,F%
\overline{Z}\right] \right) .
\end{equation*}
We then use cyclicity to show that 
\begin{eqnarray*}
F\left[ FZ,F\overline{Z}\right] &=&F^{4}+F^{2}\left[ F,\overline{Z}\right]
Z+F^{2}\left[ Z,F\right] \overline{Z} \\
&=&F^{4}+ZF^{2}\left[ F,\overline{Z}\right] +\overline{Z}F^{2}\left[ Z,F%
\right] \\
&=&F^{4}+F^{2}\overline{Z}\left[ Z,F\right] +\overline{Z}F^{2}\left[ Z,F%
\right]
\end{eqnarray*}
and therefore that 
\begin{equation}
w=3F^{4}+\frac{3}{2}\left( F^{2}\overline{Z}\left[ Z,F\right] +\overline{Z}%
F^{2}\left[ Z,F\right] \right) .  \label{eqfine}
\end{equation}
Now, since 
\begin{equation*}
F^{2}\overline{Z}\left[ Z,F\right] =[F^{2}\overline{Z},Z]F=-F^{4}-\overline{Z%
}F^{2}\left[ Z,F\right] -F\overline{Z}F\left[ Z,F\right]
\end{equation*}
and since 
\begin{eqnarray*}
-F\overline{Z}F\left[ Z,F\right] &=&FDF\overline{D}F-\overline{Z}F^{2}\left[
Z,F\right] \\
&=&-F\overline{D}FDF-F^{2}\overline{Z}\left[ Z,F\right] \\
&=&\frac{1}{2}\left( F[DF,\overline{D}F]-\overline{Z}F^{2}\left[ Z,F\right]
-F^{2}\overline{Z}\left[ Z,F\right] \right) ,
\end{eqnarray*}
we obtain that 
\begin{equation*}
\frac{3}{2}\left( F^{2}\overline{Z}\left[ Z,F\right] +\overline{Z}F^{2}\left[
Z,F\right] \right) =-F^{4}+\frac{1}{2}F[DF,\overline{D}F].
\end{equation*}
The above fact, together with (\ref{eqfine}), concludes the proof of the
lemma. $\square $\pagebreak

\end{document}